\documentclass[twocolumn, journal]{IEEEtran}
\usepackage{ifpdf}
\usepackage{cite}
\usepackage{bbm}
\ifCLASSINFOpdf
\usepackage{graphicx}
\usepackage{subfigure}
\usepackage{epstopdf}
\usepackage{xcolor}
\else
\fi
\usepackage{float}
\usepackage{amsmath}
\usepackage{amssymb}
\usepackage{amsthm}
\usepackage{mathtools}
\DeclareMathOperator*{\argmin}{arg\,min}
\newtheorem{lemma}{Lemma}
\newtheorem{prop}{Proposition}
\newtheorem{thm}{Theorem}

\newtheorem{assume}{Assumption}
\usepackage{algorithm}
\usepackage{algpseudocode}

\allowdisplaybreaks
\usepackage{array}
\usepackage{stfloats}
\usepackage{tikz}

\usepackage[shortlabels]{enumitem}
\usepackage{comment}
\usepackage{multirow}
\usepackage{cleveref}
\usepackage{booktabs}

\begin{document}
\title{Federated Learning Enhanced by Feature Reconstruction for Semantic Communication Module Updates of Agents}

%
%
%

\author{
	\IEEEauthorblockN{Yoon Huh},~\IEEEmembership{Graduate Student Member,~IEEE},
    \IEEEauthorblockN{Bumjun Kim},~\IEEEmembership{Graduate Student Member,~IEEE},
	and
	\IEEEauthorblockN{Wan Choi},~\IEEEmembership{Fellow,~IEEE}
	\thanks{A part of this paper has been presented in IEEE Global Communications Conference (GLOBECOM), Taipei, Taiwan, Dec. 2025.} 
	\thanks{Y.~Huh, B.~Kim and W.~Choi are with the Department of Electrical and Computer Engineering, Seoul National University (SNU), and the Institute of New Media and Communications, SNU, Seoul 08826, Korea. (e-mail: \{mnihy621, eithank96, wanchoi\}@snu.ac.kr) (\emph{Corresponding Authors: Wan Choi}) }
        \vspace{-3mm}

}

%

\maketitle

\begin{abstract}
Recent advancements in semantic communication have primarily focused on image transmission, where neural network-based joint source-channel coding modules play a central role. However, such systems often experience semantic communication errors due to mismatched knowledge bases between agents and performance degradation from outdated models, necessitating regular model updates. To address these challenges in vector quantization (VQ)-based image semantic communication systems, we propose FedSFR, a novel federated learning framework that incorporates semantic feature reconstruction (FR). FedSFR introduces an FR step at the parameter server and allows a subset of clients to transmit compact feature vectors in lieu of sending full local model updates, thereby improving training stability and communication efficiency. To enable effective FR learning, we design a loss function tailored for VQ-based image semantic communication and demonstrate its validity as a surrogate for image reconstruction error. We further establish a rigorous convergence analysis of FedSFR. Experimental results on two benchmark datasets validate the superiority of FedSFR over existing baselines, especially in capacity-constrained settings, confirming both its effectiveness and robustness.
\end{abstract}

\begin{IEEEkeywords}
Semantic communication, federated learning, vector quantization, autoencoder, feature reconstruction.
\end{IEEEkeywords}

\IEEEpeerreviewmaketitle
\vspace{-3mm}
\section{Introduction}
To support more spectrally efficient communications in sixth-generation (6G) networks, where the paradigm is shifting towards task-oriented communication \cite{shi2023task}, semantic communication \cite{luo2022semantic} has emerged as a key enabling technology. By prioritizing the transmission of information directly relevant to the intended task, semantic communication significantly reduces communication overhead and flexibly accommodates diverse types of source data. Among its various applications, a notable example is image reconstruction, referred to as image semantic communication, which is our primary focus.

Semantic communication typically leverages joint source-channel coding (JSCC) techniques \cite{huh2025universal}. In the context of image semantic communication, JSCC is commonly implemented using neural networks (NNs) with an autoencoder architecture, consisting of an encoder and a decoder. This design enables the joint optimization of source and channel coding, effectively reducing transmission redundancy and improving communication efficiency. In practice, JSCC models used by semantic communication users, referred to as agents,  are trained based on their individual knowledge bases (KBs), captured through their respective local datasets \cite{luo2022semantic}. However, in multi-user environments, synchronizing JSCC models requires aggregating information from all KBs. Transmitting these KBs to a server poses significant privacy concerns, as they may contain sensitive information. Furthermore, over time, the underlying data distribution of each KB may gradually shift.\footnote{While continual learning techniques can address such distribution shifts, e.g., to mitigate catastrophic forgetting, this paper focuses on developing an FL framework rather than addressing continual learning challenges.} These shifts can lead to model obsolescence and performance degradation, thereby necessitating periodic model updates.

Federated learning (FL) has emerged as an effective solution to these challenges \cite{xing2023multi, nguyen2024efficient, sun2024federated, lu2024efficient, xu2024federated}. FL enables the periodic refinement of NN-based JSCC models by aggregating knowledge from multiple decentralized sources, all while preserving data privacy. In this framework, semantic communication agents act as FL clients, each equipped with a JSCC encoder and decoder, serving as the transmitter and receiver, respectively, and together constituting the local model for FL. During training, a parameter server (PS) is responsible for aggregating the local JSCC model updates. However, once training is complete, semantic communications are carried out solely by the clients, and the PS is no longer involved.

Nonetheless, FL encounters notable challenges in resource-constrained communication environments \cite{kairouz2021advances, kim2026communication, Hong2023Base}. Aggregating updates from numerous clients can incur substantial communication overhead, as each client typically transmits its entire local model to the PS. To mitigate this, the sparsification method \cite{stich2018sparsified} was introduced, whereby only the most critical components of local models are transmitted. In addition, error feedback mechanisms \cite{karimireddy2019error}, which store the compression-induced errors and incorporate them into future updates, were employed to compensate for information loss caused by sparsification. These methods significantly improve communication efficiency under constrained network conditions.

Several studies have explored FL tailored to semantic communication. \cite{xing2023multi} and \cite{nguyen2024efficient} refined the aggregation of local JSCC models by adjusting weights based on additional local training losses reported by clients. In parallel, \cite{sun2024federated, lu2024efficient, xu2024federated} proposed FL approaches incorporating knowledge distillation (KD). However, these methods adopt a general FL perspective and overlook the distinct characteristics of image semantic communication, where the JSCC encoder and decoder naturally form an autoencoder with inverse functionality. Furthermore, they fail to consider capacity-constrained FL scenarios, particularly in the context of optimizing vector quantization (VQ)-based image semantic communication systems under digital transmission settings. In such systems, before the transmission, the encoder-generated feature vectors are quantized using a VQ codebook to obtain codeword indices, which are then transmitted.

To address this specific setting, we propose a FL framework with feature reconstruction (FR), termed FedSFR. In FedSFR, each participating client dynamically selects its transmission strategy based on channel quality: clients with favorable channels upload memory-compressed local model updates, while those with poor channels transmit semantically compact feature vectors generated by their locally updated JSCC encoders to the PS. At the PS, the received local models are first aggregated and then further refined through an additional FR process. Specifically, the received feature vectors are sequentially passed through the decoder and then through the encoder, reversing the client-side processing order during local training, where source images are first processed by the encoder and then reconstructed by the decoder. This reversed flow exploits the inherent autoencoder structure of the JSCC model to improve the consistency of feature representations while maintaining the benefits of model aggregation. By offering two strategies, FedSFR enables more efficient utilization of communication resources and achieves a stable and effective FL process compared to conventional methods.

\vspace{-3mm}
\subsection{Related Works}
\subsubsection{Aggregation Weight Design Approach}
Unlike the FedAvg algorithm \cite{mcmahan2017communication}, which assigns the aggregation weights as $p_k^{\scriptstyle(t)}=\frac{|\mathcal{D}_k|}{\sum_{i=1}^K{|\mathcal{D}_i|}}$ based on the size of each client's local dataset $|\mathcal{D}_k|$, \cite{xing2023multi} and \cite{nguyen2024efficient} designed $p_k^{\scriptstyle(t)}$ by utilizing the local training loss $l_k^{\scriptstyle(t)}$ reported from the client $k$ at global iteration $t$. They experimentally demonstrated that this loss-aware weighting can achieve better image transmission quality compared to FedAvg. Specifically, the authors of \cite{xing2023multi} proposed setting $p_k^{\scriptstyle(t)} = \frac{\exp(\tilde{l}_k^{\scriptstyle(t)})}{\sum_{i = 1}^K \exp(\tilde{l}_i^{\scriptstyle(t)})}$, where $\tilde{l}_k^{\scriptstyle(t)} = \frac{l_k^{\scriptstyle(t)} - \min_k l_k^{\scriptstyle(t)}}{\max_k l_k^{\scriptstyle(t)}- \min_k l_k^{\scriptstyle(t)}}$. This Softmax-based approach assigns higher weights to local models with larger loss values, under the assumption that such models provide more informative gradients for the global update. In contrast, \cite{nguyen2024efficient} proposed setting $p_k^{\scriptstyle(t)} = \frac{1}{K - 1}\frac{(\sum_{i = 1}^K l_i^{\scriptstyle(t)}) - l_k^{\scriptstyle(t)}}{\sum_{i = 1}^K l_i^{\scriptstyle(t)}}$,  which gives greater weight to clients with smaller losses, based on the intuition that models closer to the optimum should have a greater influence on the global model. However, both methods primarily aim to enhance model performance through aggregation strategies and do not address transmission constraints or the optimization of VQ-based image semantic communication systems under digital communication settings.

\subsubsection{Knowledge Distillation Aided Approach}
The authors of \cite{sun2024federated} proposed an FL framework that optimizes JSCC models for image classification, wherein some clients upload local models while others transmit encoder output feature vectors along with corresponding class labels. The PS updates the global JSCC decoder via KD using the received features and labels, followed by aggregation with the local models. However, this results in an imbalanced update between the JSCC encoder and decoder. Furthermore, in image reconstruction tasks, class labels can directly reveal the original image content, raising significant privacy concerns and making this approach unsuitable for image semantic communication.  

Similarly, \cite{lu2024efficient}, which also targets classification tasks, considered transmission of a compact JSCC model that carries either local updates during uplink or the global model during downlink. This compact model is smaller in size than the local JSCC model, thereby reducing communication overhead. Clients simultaneously update both their local and compact models using KD by minimizing the output discrepancy between the two models, while the PS aggregates the compact models received from clients. In contrast, \cite{xu2024federated} extended FL to image reconstruction by exchanging encoder output features and intermediate decoder features between the clients and PS. For both the clients and PS, KD is then applied to update only the model components lying between the two transmitted feature types, while the remaining parts are trained solely at each client. Nevertheless, both \cite{lu2024efficient} and \cite{xu2024federated} do not perform full aggregation of the local JSCC models, leading to inconsistent feature representations across clients and ultimately limiting the overall performance compared to conventional FL.

\vspace{-3mm}
\subsection{Contributions}
The contributions of this article are summarized as:
\begin{itemize}
    \item We propose a novel FL framework, FedSFR, designed for image semantic communication module update in capacity-constrained environments. The proposed scheme utilizes the same fixed VQ codebook of the semantic communication module during the JSCC model update by FedSFR.
    \item A dedicated FR learning process is introduced, tailored to VQ-based image semantic communication. To provide deeper insight, we explain with detailed mathematical derivations. How FR learning benefits image reconstruction tasks from two complementary perspectives: first-order Taylor expansion and Lipschitz continuity.
    \item We rigorously establish the convergence properties of FedSFR under standard FL assumptions, deriving explicit bounds on convergence rate and characterizing how FR impacts optimization dynamics.
    \item Extensive experiments on both low-resolution and high-resolution datasets, including heterogeneous local data distributions, are conducted to validate the performance of FedSFR. The results demonstrate notable improvements in training stability and convergence speed compared to existing baselines, confirming both its practical effectiveness and theoretical robustness.
\end{itemize}

Recently, \cite{huh2025feature}, our prior work, introduced an FR-aided FL framework for optimizing JSCC models using only the encoder and decoder, without considering the VQ codebook. While the focus was primarily on analog JSCC settings, the proposed framework was not inherently limited to analog communication. In contrast, this paper targets digital communication and proposes a dedicated framework specifically designed for VQ-based image semantic communication. Additionally, we provide a rigorous convergence analysis alongside an intuitive mathematical explanation of the benefits of FR learning.


\vspace{-3mm}
\section{System Model}\label{sec:system model}
Before delving into the proposed structure, we first present a basic framework for semantic communication and FL for image transmission under capacity-constrained digital communication scenarios. In this framework, semantic communication occurs after the FL process, where the FL clients act as agents in the subsequent semantic communication phase.

\begin{figure}[!t]
    \centering
    \includegraphics[width=0.95\linewidth]{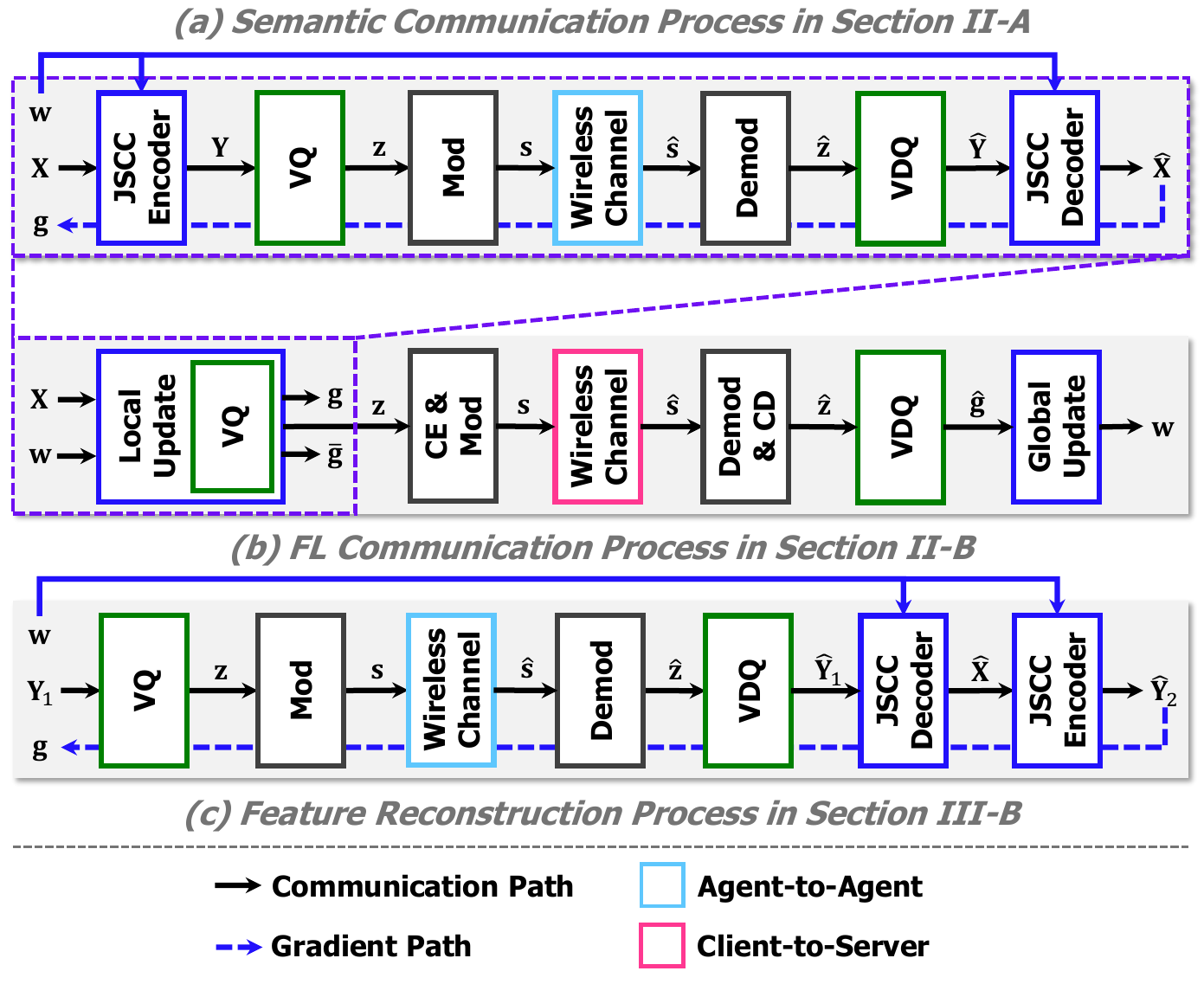}
    \vspace{-4mm}
    \caption{An illustration of (a) the semantic communication process, (b) the FL communication process, and (c) the proposed FR process, where all VQ and DVQ blocks share the same fixed VQ codebook. Semantic encoder and decoder shown in (a) are trained by  either FL in (b) or FR in (c), and then used for semantic communications. }
    \vspace{-6mm}
    \label{fig:System model}
\end{figure}
\vspace{-3mm}
\subsection{Image Semantic Communication}\label{subsec:image semantic communication}
Consider a VQ-based digital semantic communication system \cite{tung2022deepjscc} designed for image transmission, which employs an NN-based JSCC encoder, decoder, and a fixed two-dimensional VQ codebook\footnote{Using a fixed two-dimensional VQ codebook may slightly limit the maximum achievable performance at high SNR but enables to share the same VQ codebook for both semantic communication and federated learning; however, this loss is typically negligible. Notably, thanks to the inherently small VQ error, the performance degrades smoothly as the SNR decreases.} $\mathbf{C} = [\mathbf{c}_1, \mathbf{c}_2, \dots, \mathbf{c}_M]^\mathsf{T}\in\mathbb{R}^{M\times 2}$, where $M$ is a digital modulation order, e.g., $M = 256$ for $256$-QAM. Each codeword is associated with a modulation symbol in the constellation set $\mathbb{S} = \{\mathbf{s}_1, \mathbf{s}_2, \dots, \mathbf{s}_M\}\in\mathbb{R}^{2\times M}$, with normalized power $\mathbb{E}[\|\mathbf{s}_i\|_2^2] = P$. Since both the VQ codewords and modulation symbols are two-dimensional vectors, we simply set $\mathbf{c}_i = \mathbf{s}_i$ for all $i\in[1{:}M]$, and the codebook $\mathbf{C}$ is fixed during the model training and image transmission.

The communication process in Fig. \ref{fig:System model}(a) is specified as follows. At the transmitter side, a source image $\mathbf{X} \in \mathbb{R}^{C \times H \times W}$, where $C$, $H$, and $W$ denote the number of channels, height, and width of the image, respectively, is encoded into a set of feature vectors $\mathbf{Y} = [\mathbf{y}_1, \mathbf{y}_2, \dots, \mathbf{y}_N]^\mathsf{T}\in\mathbb{R}^{N\times 2}$ via an encoder $f_{\boldsymbol{\theta}}$ parameterized by $\boldsymbol{\theta}$. Here, $N$ represents the number of feature vectors. This encoding process is mathematically expressed as $\mathbf{Y} = f_{\boldsymbol{\theta}}(\mathbf{X})$. Unlike \cite{tung2022deepjscc}, the last layer of our encoder adopts the form $(1 + \mu)\cdot\mathsf{max}(\mathbf{C})\cdot\mathsf{tanh}(\cdot)$, where $\mathsf{tanh}(\cdot)$ maps its input into $[-1, 1]$, $\mathsf{max}(\mathbf{C})$ denotes the maximum absolute value among all entries of $\mathbf{C}$, and $\mu$ is a margin parameter. This design facilitates stable training and accelerates convergence.

Subsequently, a codeword index vector $\mathbf{z} = [z_1, z_2, \dots, z_N]^\mathsf{T}\in[1{:}M]^N$ is generated by quantizing $\mathbf{Y}$ using the fixed codebook $\mathbf{C}$ with the minimum distance criterion. Specifically, the $i$-th feature vector $\mathbf{y}_i$ is mapped to the codeword index $z_i$ as $z_i = \argmin_{j} \|\mathbf{y}_i - \mathbf{c}_j\|_2^2$ for all $i\in[1{:}N]$. The index vector $\mathbf{z}$ is then modulated into a symbol sequence $\mathbf{s} = [\mathbf{s}_{z_1}, \mathbf{s}_{z_2}, \dots, \mathbf{s}_{z_N}]\in\mathbb{S}^N$ and transmitted through a wireless channel. So far, the data flow at the transmitter side is summarized as $\mathbf{X} \rightarrow \mathbf{Y} \rightarrow \mathbf{z} \rightarrow \mathbf{s}$.

At the receiver side, the received symbol sequence is expressed as $\hat{\mathbf{s}} = h\mathbf{s} + \mathbf{n}$, where $h$ denotes the channel gain\footnote{Depending on the channel coherence time, $h$ may either remain unchanged or vary during a single image transmission. For notational convenience, however, we assume that $h$ remains constant.} and $\mathbf{n} \sim \mathcal{CN}(\mathbf{0}_N, \sigma^2\mathbf{I}_N)$ represents additive white Gaussian noise (AWGN). When $h = 1$, the channel reduces to an AWGN channel, whereas $h \sim \mathcal{CN}(0, 1)$ corresponds to a Rayleigh fading channel. The receiver is assumed to have perfect knowledge of the channel gain. Accordingly, using matched filtering, the symbol sequence $\bar{\mathbf{s}} = [\mathbf{s}_{\hat{z}_1}, \mathbf{s}_{\hat{z}_2}, \dots, \mathbf{s}_{\hat{z}_N}]\in\mathbb{S}^N$ is detected from $\frac{h^*}{|h|^2}\hat{\mathbf{s}} = \mathbf{s} + \frac{h^*}{|h|^2}\mathbf{n}$. The average signal-to-noise ratio (SNR) is defined as $P/\sigma^2$. Then, the detected index vector $\hat{\mathbf{z}} = [\hat{z}_1, \hat{z}_2, \dots, \hat{z}_N]^\mathsf{T}\in[1{:}M]^N$ is demapped into the received feature vectors $\hat{\mathbf{Y}} = [\mathbf{c}_{\hat{z}_1}, \mathbf{c}_{\hat{z}_2}, \dots, \mathbf{c}_{\hat{z}_N}]^\mathsf{T}\in\mathbb{R}^{N\times 2}$ using the shared codebook $\mathbf{C}$ as in the transmitter. Finally, the reconstructed image $\hat{\mathbf{X}}\in\mathbb{R}^{C\times H\times W}$ is obtained from $\hat{\mathbf{Y}}$ via a decoder $f^{-1}_{\boldsymbol{\phi}}$, parameterized by $\boldsymbol{\phi}$. This decoding process is represented as $\hat{\mathbf{X}} = f^{-1}_{\boldsymbol{\phi}}(\hat{\mathbf{Y}})$. To wrap up the process, the data flow at the receiver side can be represented as $\hat{\mathbf{s}} \rightarrow \bar{\mathbf{s}} \rightarrow \hat{\mathbf{z}} \rightarrow \hat{\mathbf{Y}} \rightarrow \hat{\mathbf{X}}$.

The learnable parameters in the semantic communication module are denoted as $\boldsymbol{w} = \{\boldsymbol{\theta}, \boldsymbol{\phi}\} \in \mathbb{R}^{D}$, where $D$ represents the total number of parameters to be optimized. The parameter $\boldsymbol{w}$ is trained utilizing the loss function in \cite{tung2022deepjscc}, given by
\begin{align}\label{eq:image reconstruction loss}
     l_c(\boldsymbol{w}; \mathbf{X}) = \mathsf{MSE}(\hat{\mathbf{X}}, \mathbf{X}) + \alpha\cdot\mathsf{KLD}(\hat{P}(\mathbf{C})\,||\,\mathcal{U}[1{:}M]),
\end{align}
where $\mathsf{MSE}(\mathbf{A}, \mathbf{B})$ is the mean squared error (MSE) between $\mathbf{A}$ and $\mathbf{B}$, $\mathsf{KLD}(P||Q)$ represents the Kullback-Leibler (KL) divergence between the distribution $P$ and $Q$, and $\alpha$ is a hyperparameter balancing the reconstruction and regularization terms. Here, $\hat{P}(\mathbf{C})$ denotes the empirical usage distribution of the codewords in $\mathbf{C}$, while $\mathcal{U}[1{:}M]$ denotes the uniform distribution over the $M$ codewords. The first term in $l_c(\boldsymbol{w}; \mathbf{X})$ encourages the encoder–decoder pair to accurately reconstruct the image, taking into account the VQ operation and the AWGN channel. The second term regularizes the codeword usage to approach a uniform distribution, thereby promoting balanced exploration of all codewords during training. The global objective function at the PS in FL is defined as $F(\boldsymbol{w}) = \frac{1}{|\mathcal{D}|}\sum_{\mathbf{X}\in\mathcal{D}}l_c(\boldsymbol{w}; \mathbf{X})$, where $\mathcal{D}$ denotes the source image dataset. In this paper, we assume that all objective functions are non-convex, as discussed in \cite{karimireddy2019error}.

\vspace{-3mm}
\subsection{Federated Learning}\label{subsec:federated learning}
Consider a digital FL using the FedAvg algorithm \cite{mcmahan2017communication}, where the PS collaboratively optimizes the global model $\boldsymbol{w}$, including the JSCC encoder $f_{\boldsymbol{\theta}}$ and decoder $f^{-1}_{\boldsymbol{\phi}}$, with $K$ clients in the set $\mathcal{A}$, i.e., $|\mathcal{A}| = K$. Each client $k$ holds a local dataset $\mathcal{D}_k$, and the global dataset is defined as the union of all local datasets, i.e., $\mathcal{D} = \bigcup_{k \in \mathcal{A}} \mathcal{D}_k$. For each client, the local objective function is expressed as $F_k(\boldsymbol{w}) = \frac{1}{|\mathcal{D}_k|}\sum_{\mathbf{X}\in\mathcal{D}_k}l_c(\boldsymbol{w}; \mathbf{X})$. Then, the global objective function is formulated as $F(\boldsymbol{w}) = \sum_{k\in\mathcal{A}}p_k F_k(\boldsymbol{w})$, where $p_k = |\mathcal{D}_k| / |\mathcal{D}|$ denotes the proportion of the global dataset associated with client $k$.

FL process is carried out iteratively through local and global update steps. During the local update process, each client downloads the global model, denoted as $\boldsymbol{w}_k^{\scriptstyle(t, 0)} = \boldsymbol{w}^{\scriptstyle(t)} \in \mathbb{R}^{D}$, from the PS at global iteration $t \in \{0, \dots, T - 1\}$. Subsequently, the client updates $\boldsymbol{w}_k^{\scriptstyle(t, 0)}$ locally using its dataset $\mathcal{D}_k$ by employing a mini-batch stochastic gradient descent (SGD) algorithm. The local update rule is given by
\begin{align}\label{eq:local iteration}
    \boldsymbol{w}_k^{\scriptstyle(t, e + 1)} = \boldsymbol{w}_k^{\scriptstyle(t, e)} - \eta_c^{\scriptstyle(t)} \nabla F_k^{\scriptstyle(t, e)}(\boldsymbol{w}_k^{\scriptstyle(t, e)}), 
\end{align}
where $e \in \{0, \dots, E_c - 1\}$ is the local iteration index, $\eta_c^{\scriptstyle(t)}$ denotes the local learning rate at global iteration $t$, and $E_c$ is the total number of local iterations. At each local iteration $e$, the local gradient vector is computed as
\begin{align} \nabla F_k^{\scriptstyle(t, e)}(\boldsymbol{w}_k^{\scriptstyle(t, e)}) = \frac{1}{|\mathcal{D}_k^{\scriptstyle(t, e)}|} \sum_{\mathbf{X} \in \mathcal{D}_k^{\scriptstyle(t, e)}} \nabla l_c(\boldsymbol{w}_k^{\scriptstyle(t, e)}; \mathbf{X}), 
\end{align}
where $\mathcal{D}_k^{\scriptstyle(t, e)}$ is a mini-batch sampled from $\mathcal{D}_k$ and $l_c(\boldsymbol{w}; \mathbf{X})$ is computed using $\hat{\mathbf{X}}$ obtained by emulating the wireless channel for semantic transmission between agents as in Fig. \ref{fig:System model}(a).

To meet the uplink capacity constraint between the $k$-th client and the PS, the client compresses the local update information $\mathbf{g}_k^{\scriptstyle(t)}\in\mathbb{R}^{D}$ into a compressed version $\bar{\mathbf{g}}_k^{\scriptstyle(t)}\in\mathbb{R}^{D}$ using a top-$S$ sparsification technique\footnote{The top-$S$ sparsification algorithm selects the $S$ largest values in the vector by ranking elements based on their magnitudes. In this paper, we apply sparsification on a per-layer basis in NN. Note that the indices corresponding to the top-$S$ positions are also transmitted to the PS. However, since the communication overhead for these indices is negligible compared to the overall model size $D$, we omit it from our analysis.} combined with an error-feedback strategy\footnote{Error-feedback is widely used with top-$S$ sparsification because sparsification causes biased updates. Accumulating and re-injecting these residuals effectively restores lost local update information and significantly improves convergence performance \cite{karimireddy2019error}.}. The local update information at the $k$-th client is defined as
\begin{align}\label{eq:local update information}
    \mathbf{g}_k^{\scriptstyle(t)} = \mathbf{m}_k^{\scriptstyle(t)} + \eta_c^{\scriptstyle(t)}\sum\nolimits_{e = 0}^{E_c - 1}\nabla F_k^{\scriptstyle(t, e)}(\boldsymbol{w}_k^{\scriptstyle(t, e)}),
\end{align}
and its compressed version is represented as
\begin{align}\label{eq:compression}
    \bar{\mathbf{g}}_k^{\scriptstyle(t)} = \mathsf{Compress}(\mathbf{g}_k^{\scriptstyle(t)}),
\end{align}
where $\mathsf{Compress}$ denotes the compression method that sequentially performs sparsification, quantization, and dequantization. Using the same VQ codebook $\mathbf{C}$ in Section \ref{subsec:image semantic communication}, the index vector $\mathbf{z}_k^{\scriptstyle(t)}$ is generated after the quantization process described in Section \ref{subsubsec:gradient quantization}, and then dequantized into $\bar{\mathbf{g}}_k^{\scriptstyle(t)}$. To mitigate the effects of compression error in the subsequent global iteration, the error memory feedback strategy is employed. The error memory $\mathbf{m}_k^{\scriptstyle(t)}\in\mathbb{R}^{D}$ at the $k$-th client is denoted as
\begin{align}\label{eq:error memory update}
    \mathbf{m}_k^{\scriptstyle(t + 1)} &= \mathbf{g}_k^{\scriptstyle(t)} - \bar{\mathbf{g}}_k^{\scriptstyle(t)}\\
    &= \mathbf{m}_k^{\scriptstyle(t)} + \eta_c^{\scriptstyle(t)}\sum\nolimits_{e = 0}^{E_c - 1}\nabla F_k^{\scriptstyle(t, e)}(\boldsymbol{w}_k^{\scriptstyle(t, e)}) - \bar{\mathbf{g}}_k^{\scriptstyle(t)}.
\end{align}

Consider a subset of participating clients, $\mathcal{A}_m^{\scriptstyle(t)}\subset\mathcal{A}$, at global iteration $t$, where $|\mathcal{A}_m^{\scriptstyle(t)}| = K_m$. Note that the subscript `$m$' stands for local \underline{m}odel. As shown in Fig. \ref{fig:System model}(b), these clients transmit symbol sequences $\mathbf{s}_k^{\scriptstyle(t)}$, obtained by modulating bit sequences generated from the channel-encoded version of $\mathbf{z}_k^{\scriptstyle(t)}$, via uplink transmission. Similar to semantic transmission, the uplink channel follows either an AWGN or a Rayleigh fading model, i.e., $\hat{\mathbf{s}}_k^{\scriptstyle(t)} = h_k^{\scriptstyle(t)}\mathbf{s}_k^{\scriptstyle(t)} + \mathbf{n}_k^{\scriptstyle(t)}$. The total number of bits transmitted by client $k$ within the latency constraint $T$ is defined as $B_k = \lfloor C_k TR\rfloor$, where $C_k$ and $R$ denote the uplink channel capacity for client $k$ and the channel coding rate, respectively, ensuring error-free transmission of $\bar{\mathbf{g}}_k^{\scriptstyle(t)}$ according to the channel capacity. Given the codebook structure $\mathbf{C}\in\mathbb{R}^{M\times 2}$, the corresponding sparsification level is determined as $S_k = \lfloor 2B_k/\log_2 M\rfloor$.

 The PS can reliably recover the index vector $\hat{\mathbf{z}}_k^{\scriptstyle(t)} = \mathbf{z}_k^{\scriptstyle(t)}$ through channel equalization, symbol detection, and channel decoding of $\hat{\mathbf{s}}_k^{\scriptstyle(t)}$, and then reconstructs $\hat{\mathbf{g}}_k^{\scriptstyle(t)} = \bar{\mathbf{g}}_k^{\scriptstyle(t)}$. Subsequently, the global update process is carried out through weighted averaging \cite{li2019convergence}, given by
\begin{align}\label{eq:global update}
    \boldsymbol{w}^{\scriptstyle(t + 1)} = \boldsymbol{w}^{\scriptstyle(t)} - \frac{K}{K_m}\sum_{k\in\mathcal{A}_m^{\scriptstyle(t)}}p_k \bar{\mathbf{g}}_k^{\scriptstyle(t)}.
\end{align}
Finally, the PS broadcasts the updated global model $\boldsymbol{w}^{\scriptstyle(t + 1)}$ to all clients via the downlink transmission.

\vspace{-3mm}
\section{Proposed FedSFR Algorithm}\label{sec:proposed fedsfr algorithm}
Transmitting full local update information at every iteration imposes significant communication overhead. To mitigate this, we propose FedSFR, which leverages feature vectors instead of local update information for clients experiencing poor channel conditions. The proposed FL algorithm with FR is detailed in Section \ref{subsec:algorithm description}. Subsequently, we introduce the  FR loss function, discuss its design rationale, and explain how FR learning improves image transmission performance in Sections \ref{subsec:feature reconstruction learning} and \ref{subsec:from feature reconstruction to image reconstruction}, respectively. For clarity, the key notations used throughout the framework are summarized in Table \ref{tab:Notations about FedSFR at global iteration t}.

\begin{table}[t!]
\caption{Notations about FedSFR at global iteration $t$.}
\vspace{-2mm}
\label{tab:Notations about FedSFR at global iteration t}
\centering
\begin{tabular}{cl}
\hline
\textbf{Notation}                                    & \textbf{Description}                                                                                     \\ \hline
$N$                                                  & Number of feature vectors encoded from a single image                                                    \\
$D$                                                  & Number of JSCC model parameters                                                                          \\
$M$                                                  & Codebook size                                                                                            \\
$\mu$                                                & Margin parameter for JSCC encoder output values                                                          \\
$\boldsymbol{w}^{\scriptstyle(t)}$                   & Global model to be downloaded at the clients                                                             \\
$\boldsymbol{w}^{\scriptstyle(t + \frac{1}{2})}$     & Global model to be updated at the PS                                                                     \\
$\boldsymbol{w}_k^{\scriptstyle(t,e)}$               & Local model of client $k$ at local iteration $e$                                                         \\
$\boldsymbol{w}_s^{\scriptstyle(t + \frac{1}{2},e)}$ & Server model at server iteration $e$                                                                     \\
$\mathbf{g}_k^{\scriptstyle(t)}$                     & Local update information at client $k$                                                                   \\
$\bar{\mathbf{g}}_k^{\scriptstyle(t)}$               & Compressed local update information at client $k$                                                        \\
$\mathbf{m}_k^{\scriptstyle(t)}$                     & Error memory before quantization at client $k$                                                           \\
$\mathcal{P}_k$                                      & Shared public dataset at client $k$                                                                      \\
$\mathcal{Y}_k^{\scriptstyle(t)}$                    & Set of feature vectors after local update at client $k$                                                  \\
$K$ ($\mathcal{A}$)                                  & Number (Set) of total clients                                                                            \\
$K_m$ ($\mathcal{A}_m^{\scriptstyle(t)}$)            & Number (Set) of clients sending $\bar{\mathbf{g}}_k^{\scriptstyle(t)}$                                   \\
$K_o$ ($\mathcal{A}_o^{\scriptstyle(t)}$)            & Number (Set) of clients sending $\mathcal{Y}_k^{\scriptstyle(t)}$                                        \\
$S_m$ / $S_o$                                        & Sparsification level of clients in $\mathcal{A}_m^{\scriptstyle(t)}$ / $\mathcal{A}_o^{\scriptstyle(t)}$ \\
$T$                                                  & Number of global iterations                                                                              \\
$E_c$                                                & Number of local iterations                                                                               \\
$E_s$                                                & Number of server iterations                                                                              \\ \hline
\end{tabular}
\vspace{-5mm}
\end{table}

\vspace{-3mm}
\subsection{Algorithm Description}\label{subsec:algorithm description}
We propose that clients experiencing poor channel conditions transmit compact feature vectors $\mathbf{Y}$, generated from the most recent updates of their local JSCC encoder, instead of directly transmitting local update information $\mathbf{g}_k^{\scriptstyle(t)}$. This approach exploits the fact that the model size $D$ (i.e., $\boldsymbol{w}$) is typically much larger than the feature vector size $2N$ (i.e., $\mathbf{Y}$) in image semantic communication, i.e., $D \gg 2N$. In reconstruction tasks, feature vectors are produced at an intermediate stage of the model, specifically after the encoder. Given that the encoder and decoder are designed to function as approximate inverses, these vectors inherently encapsulate both model and data-specific information, offering a compact yet highly informative representation. Transmitting this compressed representation may appear conceptually similar to federated KD in classification tasks, where client-side logit vectors convey local model knowledge to the server. In federated KD, knowledge transfer is achieved by aligning logits generated from the same input image during local update, using global logits as explicit supervision signals for the client logits. In contrast, our method introduces FR, where the server reconstructs transmitted feature vectors, without directly using the original images as input.

\begin{figure}[!t]
    \centering
    \includegraphics[width=1\linewidth]{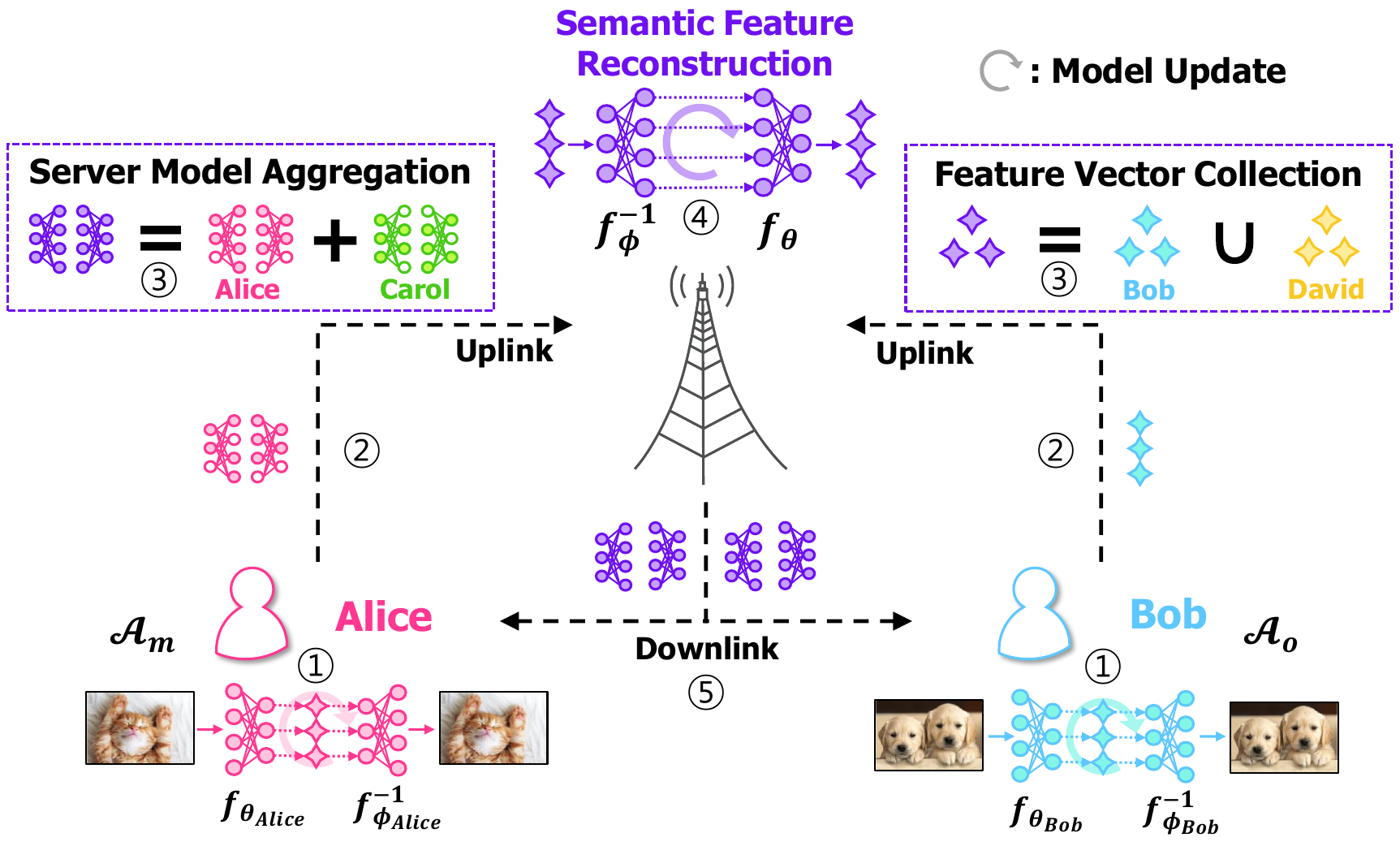}
    \vspace{-5mm}
    \caption{Overall procedure of FedSFR with the numbered algorithmic steps.}
    \vspace{-6mm}
    \label{fig:Overall procedure of FedSFR with the numbered algorithmic steps}
\end{figure}

As illustrated in Fig. \ref{fig:Overall procedure of FedSFR with the numbered algorithmic steps}, which outlines the five \texttt{steps} of the FedSFR algorithm, the procedure unfolds as follows. Let $\mathcal{A}_m^{\scriptstyle(t)}$ and $\mathcal{A}_o^{\scriptstyle(t)}$ denote two sets of participating clients at global iteration $t$, where the subscript `$o$' refers to clients transmitting encoder \underline{o}utput feature vectors. Assume that clients in $\mathcal{A}_m^{\scriptstyle(t)}$ experience better channel conditions than those in $\mathcal{A}_o^{\scriptstyle(t)}$, and let $|\mathcal{A}_o^{\scriptstyle(t)}| = K_o$. Following the local update process described in \eqref{eq:local iteration} (\texttt{step 1}), each client $k \in \mathcal{A}_m^{\scriptstyle(t)}$ transmits its compressed local update information $\bar{\mathbf{g}}_k^{\scriptstyle(t)}\in\mathbb{R}^D$ in accordance with \eqref{eq:local update information} and \eqref{eq:compression} (\texttt{step 2}), where a total transmission budget of  $B_k = B_m$. In contrast, each client $k \in \mathcal{A}_o^{\scriptstyle(t)}$ transmits a set $\mathcal{Y}_k^{\scriptstyle(t)}$ containing the quantized feature vectors $\bar{\mathbf{Y}}\in\mathbb{R}^{N\times 2}$ (\texttt{step 2}), with $B_k = B_o < B_m$. Here, the original feature vectors $\mathbf{Y}$ are computed using the local JSCC encoder $f_{\boldsymbol{\theta}}$ with parameters $\boldsymbol{\theta} = \boldsymbol{\theta}_k^{\scriptstyle(t, E_c)}$ and subsequently quantized into $\bar{\mathbf{Y}}$. Note that the same VQ codebook $\mathbf{C}$ is utilized for both the quantization of the top-$S$ sparsified local update information at the clients in $\mathcal{A}_m^{\scriptstyle(t)}$ and the feature quantization at the clients in $\mathcal{A}_o^{\scriptstyle(t)}$. Their procedures are described below.

\subsubsection{Quantization at $\mathcal{A}_m^{\scriptstyle(t)}$}\label{subsubsec:gradient quantization}
Let $\mathbf{v}_{k,i}^{\scriptstyle(t)}\in\mathbb{R}^2$ denote the $i$-th branch which is a two-dimensional segment from the top-$S$ sparsified $\mathbf{g}_k^{\scriptstyle(t)}$. To quantize it with $\mathbf{C}$, each branch is first normalized using the maximum absolute values of $\mathbf{C}$ and $\mathbf{g}_k^{\scriptstyle(t)}$, denoted by $\mathsf{max}(\mathbf{C})$ and $\mathsf{max}(\mathbf{g}_k^{\scriptstyle(t)})$, respectively: $\tilde{\mathbf{v}}_{k, i}^{\scriptstyle(t)} = \mathbf{v}_{k, i}^{\scriptstyle(t)}\cdot[\mathsf{max}(\mathbf{C})/\mathsf{max}(\mathbf{g}_k^{\scriptstyle(t)})]$. Here, $\mathsf{max}(\mathbf{C})$ is a fixed constant known to all clients, whereas $\mathsf{max}(\mathbf{g}_k^{\scriptstyle(t)})$ is transmitted to the PS with negligible overhead. Next, the four nearest neighboring codewords are selected based on the squared Euclidean distance $\|\tilde{\mathbf{v}}_{k, i}^{\scriptstyle(t)} - \mathbf{c}_j\|_2^2$, resulting in the set $\{\mathbf{c}_{j}, \mathbf{c}_{j + 1}, \mathbf{c}_{j + \sqrt{M}}, \mathbf{c}_{j + \sqrt{M} + 1}\}$, where these codewords form the upper-left, upper-right, lower-left, and lower-right vertices of a local square cell surrounding the normalized branch vector $\tilde{\mathbf{v}}_{k, i}^{\scriptstyle(t)}$, respectively. $\tilde{\mathbf{v}}_{k, i}^{\scriptstyle(t)}$ is then stochastically quantized into the codeword index $l$ by sampling one of the four indices:
\begin{align}
    l =
    \begin{cases}
        j & \text{with probability of}~p_\mathsf{left}\cdot p_\mathsf{up},\\
        j + 1 & \text{with probability of}~p_\mathsf{right}\cdot p_\mathsf{up},\\
        j + \sqrt{M} & \text{with probability of}~p_\mathsf{left}\cdot p_\mathsf{down},\\
        j + \sqrt{M} + 1 & \text{with probability of}~p_\mathsf{right}\cdot p_\mathsf{down},
    \end{cases}
\end{align}
with $p_\mathsf{left} = ([\mathbf{c}_{j + 1}]_1 - [\tilde{\mathbf{v}}_{k, i}^{\scriptstyle(t)}]_1)/([\mathbf{c}_{j + 1}]_1 - [\mathbf{c}_{j}]_1)$, $p_\mathsf{right} = 1 - p_\mathsf{left}$, $p_\mathsf{up} = ([\tilde{\mathbf{v}}_{k, i}^{\scriptstyle(t)}]_2 - [\mathbf{c}_{j + \sqrt{M}}]_2)/([\mathbf{c}_{j}]_2 - [\mathbf{c}_{j + \sqrt{M}}]_2)$, and $p_\mathsf{down} = 1 - p_\mathsf{up}$. This stochastic mapping ensures that the resulting quantized estimator is unbiased, i.e., $\mathbb{E}[\mathbf{c}_l] = \tilde{\mathbf{v}}_{k, i}^{\scriptstyle(t)}$.

\subsubsection{Quantization at $\mathcal{A}_o^{\scriptstyle(t)}$}\label{subsubsec:feature quantization}
Before quantization using $\mathbf{C}$, the original feature vectors $\mathbf{Y}$ produced by the JSCC encoder are first normalized by the factor $1 + \mu$. This normalization helps keep the transmitted feature values close to their original magnitudes, thereby facilitating effective FR learning at the PS. Subsequently, the same VQ procedure used in semantic communication is applied based on Euclidean distance.

The feature vector set $\mathcal{Y}_k^{\scriptstyle(t)}$ is assumed to be derived from a shared public dataset\footnote{The public dataset is pre-distributed to clients prior to the FL process, which is a widely adopted approach in distributed learning for mitigating data and model heterogeneity among participating clients \cite{zhao2018federated, chang2021cronus, itahara2021distillation, shao2024selective}.}, which is client-specific and used as part of each client’s local dataset, i.e., $\mathcal{P}_k \subset \mathcal{D}_k$, where $\mathcal{P}_k$ denotes the public dataset available to client $k$. Local model updates are computed using the full local dataset $\mathcal{D}_k$, which contains both private and public samples, ensuring that the resulting update reflects the influence of private data. Note that the transmitted feature vectors are generated from a public dataset, rather than from clients’ private data distributions, in order to inherently reduce the risk of feature inversion attacks that seek to reconstruct client data from the transmitted representations. By appropriately selecting the size of a randomly sampled subset from $\mathcal{P}_k$, the size of $\mathcal{Y}_k^{\scriptstyle(t)}$ can be adjusted to satisfy the communication budget constraint $B_o$.

Next, the PS aggregates the compressed local updates $\{\bar{\mathbf{g}}_k^{\scriptstyle(t)}\}_{k\in\mathcal{A}_o^{\scriptstyle(t)}}$ and constructs a server-side dataset from the received feature vectors as $\mathcal{D}_s^{\scriptstyle(t + \frac{1}{2})} = \bigcup_{k \in \mathcal{A}_o^{\scriptstyle(t)}} \mathcal{Y}_k^{\scriptstyle(t)}$ (\texttt{step 3}). In the proposed method, the aggregated model can be expressed by
\begin{align}\label{eq:global aggregation}
    \boldsymbol{w}^{\scriptstyle(t + \frac{1}{2})} = \boldsymbol{w}^{\scriptstyle(t)} - \frac{K}{K_m}\sum_{k\in\mathcal{A}_m^{\scriptstyle(t)}}p_k \bar{\mathbf{g}}_k^{\scriptstyle(t)},
\end{align}
where the time index $t + \frac{1}{2}$ is the difference from the time index $t + 1$ used in \eqref{eq:global update}.

By learning to reconstruct the feature vectors received from clients in $\mathcal{A}_o^{\scriptstyle(t)}$, the PS further refines the global model, initialized as $\boldsymbol{w}_s^{\scriptstyle(t + \frac{1}{2}, 0)} = \boldsymbol{w}^{\scriptstyle(t + \frac{1}{2})}$ (\texttt{step 4}). The server-side objective function for FR at global iteration $t$ is defined as $F_s^{\scriptstyle(t + \frac{1}{2})}(\boldsymbol{w}) = \frac{1}{|\mathcal{D}_s^{\scriptstyle(t + \frac{1}{2})}|}\sum_{\bar{\mathbf{Y}}\in\mathcal{D}_s^{\scriptstyle(t + \frac{1}{2})}}l_s(\boldsymbol{w}; \bar{\mathbf{Y}})$, where $l_s(\boldsymbol{w}; \bar{\mathbf{Y}})$ denotes the FR loss function, which will be described in detail in Section \ref{subsec:feature reconstruction learning}. Since the PS possesses the aggregated global model\footnote{The PS treats the entire NN model as a unified function, $f_{\boldsymbol{\theta}}\circ f^{-1}_{\boldsymbol{\phi}}$, interpreting the JSCC encoder and decoder as a sequentially connected pipeline. In contrast, clients view the encoder and decoder as distinct modules but align their update strategy to match the PS's unified perspective.}, including the learnable JSCC encoder-decoder and the fixed VQ codebook, it can directly utilize the feature vectors in $\mathcal{D}_s^{\scriptstyle(t + \frac{1}{2})}$ to perform FR. This server-side refinement is fundamentally different from local training for image reconstruction performed by the clients.

Importantly, this process enables \textit{the indirect transfer of the knowledge from the local models}  of clients in $\mathcal{A}_o^{\scriptstyle(t)}$ to the PS, without requiring the transmission of full local update information. Given the PS's significantly greater computational resources compared to the clients, the additional overhead incurred by FR-based refinement is negligible in the context of the overall FL global iteration process \cite{kairouz2021advances}. Furthermore, once the PS successfully integrates the knowledge from clients in $\mathcal{A}_o^{\scriptstyle(t)}$ through this update, the error memory is reset to $\mathbf{m}_k^{\scriptstyle(t + 1)} = \mathbf{0}_D$ for all $k \in \mathcal{A}_o^{\scriptstyle(t)}$.

The server iteration process, utilizing the same SGD algorithm employed by the clients, is described as follows:
\begin{align}\label{eq:server iteration}
    \boldsymbol{w}_s^{\scriptstyle(t + \frac{1}{2}, e + 1)} = \boldsymbol{w}_s^{\scriptstyle(t + \frac{1}{2}, e)} - \eta_s^{\scriptstyle(t)}\nabla F_s^{\scriptstyle(t + \frac{1}{2}, e)}(\boldsymbol{w}_s^{\scriptstyle(t + \frac{1}{2}, e)}),
\end{align}
where $e\in\{0, \dots, E_s - 1\}$ represents the server iteration number, $E_s$ is the total number of server iterations, and $\eta_s^{\scriptstyle(t)}$ denotes the server learning rate at global iteration $t$. At server iteration $e$, the server gradient vector is defined as
\begin{align}
    &\nabla F_s^{\scriptstyle(t + \frac{1}{2}, e)}(\boldsymbol{w}_s^{\scriptstyle(t 
    + \frac{1}{2}, e)})\nonumber\\
    &\quad\quad = \frac{1}{|\mathcal{D}_s^{\scriptstyle(t + \frac{1}{2}, e)}|}\sum_{\bar{\mathbf{Y}}\in\mathcal{D}_s^{\scriptstyle(t + \frac{1}{2} , e)}}\nabla l_s(\boldsymbol{w}_s^{\scriptstyle(t + \frac{1}{2}, e)}; \bar{\mathbf{Y}}),
\end{align}
where a mini-batch $\mathcal{D}_s^{\scriptstyle(t + \frac{1}{2}, e)}$ is sampled from $\mathcal{D}_s^{\scriptstyle(t + \frac{1}{2})}$. After completing the server update process, the PS broadcasts the updated global model $\boldsymbol{w}^{\scriptstyle(t + 1)} = \boldsymbol{w}_s^{\scriptstyle(t + \frac{1}{2}, E_s)}$ to the clients via the downlink channel (\texttt{step 5}).

\vspace{-3mm}
\subsection{Feature Reconstruction Learning}\label{subsec:feature reconstruction learning}
As illustrated in Fig. \ref{fig:System model}(c), since both the JSCC encoder and decoder are updated by FedSFR, the proposed FR employs emulated wireless channels for semantic transmission between agents during the local update process. Let the feature vectors $\mathbf{Y}_1$ be a received feature sample from the set $\mathcal{D}_s^{(t+\frac{1}{2})}$. $\mathbf{Y}_1$ is quantized and modulated into a symbol sequence, which is then transmitted through the emulated channel, i.e., $\hat{\mathbf{s}} = h\mathbf{s} + \mathbf{n}$. After channel equalization, symbol detection, and dequantization, the resulting feature vectors are denoted by $\hat{\mathbf{Y}}_1$. Note that quantization and dequantization are performed using the VQ codebook $\mathbf{C}$.

The PS then applies the JSCC decoder $f^{-1}_{\boldsymbol{\phi}}$ and encoder $f_{\boldsymbol{\theta}}$, with parameters $\boldsymbol{\theta} = \boldsymbol{\theta}_s^{\scriptstyle(t + \frac{1}{2}, e)}$ and $\boldsymbol{\phi} = \boldsymbol{\phi}_s^{\scriptstyle(t + \frac{1}{2}, e)}$, respectively, to reconstruct the feature vectors $\mathbf{Y}_2$ as $\mathbf{Y}_2 = f_{\boldsymbol{\theta}}(f^{-1}_{\boldsymbol{\phi}}(\hat{\mathbf{Y}}_1))$. In summary, the data flow for FR learning at the PS is $\mathbf{Y}_1 \rightarrow \hat{\mathbf{Y}}_1 \rightarrow \mathbf{Y}_2$.
Then, we define the FR loss function as
\begin{align}\label{eq:feature reconstruction loss}
     l_s(\boldsymbol{w}; \mathbf{Y}_1) = \mathsf{MSE}(\mathbf{Y}_2, \mathbf{Y}_1),
\end{align}
which optimizes the JSCC encoder and decoder to accurately reconstruct the feature vectors, while accounting for the effect of the wireless channel between agents by emulating it. 

\vspace{-3mm}
\subsection{From Feature Reconstruction To Image Reconstruction}\label{subsec:from feature reconstruction to image reconstruction}
We provide mathematical intuition to clarify how learning FR at the server side contributes positively to the overall goal of training JSCC encoder and decoder via FL for semantic communication. Specifically, we demonstrate that minimizing the FR error leads to improved image reconstruction quality. To support this claim, we present two complementary analytical perspectives: one based on first-order Taylor expansion, and the other leveraging the concept of Lipschitz continuity.

\subsubsection{First-Order Taylor Expansion}
The JSCC encoder $f_{\boldsymbol{\theta}}: \mathbb{R}^{CHW}\to\mathbb{R}^{Nd}$ can be locally approximated using a first-order Taylor expansion around a reference point $\hat{\mathbf{X}}$, yielding $f_{\boldsymbol{\theta}}(\mathbf{X})^\mathsf{T} \approx f_{\boldsymbol{\theta}}(\hat{\mathbf{X}})^\mathsf{T} + (\mathbf{X} - \hat{\mathbf{X}})^\mathsf{T}\mathbf{g}$ where $\mathbf{g} = \frac{\partial f_{\boldsymbol{\theta}}(\mathbf{X})}{\partial \mathbf{X}}\big|_{\mathbf{X} = \hat{\mathbf{X}}}\in\mathbb{R}^{CHW\times Nd}$ is the Jacobian of the encoder output with respect to the input image. Letting $\mathbf{Y} = f_{\boldsymbol{\theta}}(\mathbf{X})$ and $\hat{\mathbf{Y}} = f_{\boldsymbol{\theta}}(\hat{\mathbf{X}})$, we obtain $(\mathbf{Y} - \hat{\mathbf{Y}})^\mathsf{T} \approx (\mathbf{X} - \hat{\mathbf{X}})^\mathsf{T}\mathbf{g}$. Consequently, the FR error can be expressed as $||\mathbf{Y} - \hat{\mathbf{Y}}||_2^2 \approx (\mathbf{X} - \hat{\mathbf{X}})^\mathsf{T}\mathbf{g}\mathbf{g}^\mathsf{T}(\mathbf{X} - \hat{\mathbf{X}})$. Since $\mathbf{g}\mathbf{g}^\mathsf{T}$ is positive semidefinite, its eigenvalue decomposition $\mathbf{g}\mathbf{g}^\mathsf{T} = \mathbf{U}\mathbf{\Lambda}\mathbf{U}^\mathsf{T}$ leads to
\begin{align}
    ||\mathbf{Y} - \hat{\mathbf{Y}}||_2^2 \approx \sum\nolimits_{i=1}^{CHW}\lambda_i(\mathbf{u}_i^\mathsf{T}(\mathbf{X} - \hat{\mathbf{X}}))^2,
\end{align}
where $\mathbf{\Lambda} = \mathsf{diag}(\boldsymbol{\lambda})$ contains the eigenvalues and  $\mathbf{U} = [\mathbf{u}_1, \mathbf{u}_2, \dots, \mathbf{u}_{CHW}]$ contains the corresponding orthonormal eigenvectors. 
This analysis demonstrates that the FR loss $||\mathbf{Y} - \hat{\mathbf{Y}}||_2^2$ serves as a meaningful surrogate for the image reconstruction error $||\mathbf{X} - \hat{\mathbf{X}}||_2^2$. For further analytical simplicity, assume that the Jacobian $\mathbf{g}$ is independent of $(\mathbf{X} - \hat{\mathbf{X}})$ and its elements are independently and identically distributed (i.i.d.) standard Gaussian. In this case, the expected outer product becomes
$\mathbb{E}_\mathbf{X}[\mathbf{g}\mathbf{g}^\mathsf{T}] = Nd\cdot\mathbf{I}_{CHW}$, which leads to a direct linear relationship between the expected FR error and the image reconstruction error:
\begin{align}
    \mathbb{E}_\mathbf{X}[||\mathbf{Y} - \hat{\mathbf{Y}}||_2^2] \approx Nd\cdot\mathbb{E}_\mathbf{X}[||\mathbf{X} - \hat{\mathbf{X}}||_2^2].
\end{align}

\subsubsection{Lipschitz Continuity}
Define $\check{\mathbf{Y}} = \argmin_{\mathbf{Y}_0}||\hat{\mathbf{Y}} - \mathbf{Y}_0||_2^2$, where $\mathbf{Y}_0\in\{\mathbf{Y}|f_{\boldsymbol{\phi}}^{-1}(\mathbf{Y}) = \mathbf{X}\}$ and $\hat{\mathbf{Y}} = f_{\boldsymbol{\theta}}(\hat{\mathbf{X}})$.  Assuming that the decoder $f_{\boldsymbol{\phi}}^{-1}$ is Lipschitz continuous with a constant $L$, given $\mathbf{Y} = f_{\boldsymbol{\theta}}
(\mathbf{X})$ and $\hat{\mathbf{X}} = f_{\boldsymbol{\phi}}^{-1}
(\mathbf{Y})$, we have
\begin{align*}
    ||\hat{\mathbf{X}} - \mathbf{X}||_2^2 &= ||f_{\boldsymbol{\phi}}^{-1}(\mathbf{Y}) - f_{\boldsymbol{\phi}}^{-1}(\check{\mathbf{Y}})||_2^2\\
    &\leq L^2 ||\mathbf{Y} - \check{\mathbf{Y}}||_2^2\\
    &\leq 2L^2 ||\mathbf{Y} - \hat{\mathbf{Y}}||_2^2 + 2L^2 ||\hat{\mathbf{Y}} - \check{\mathbf{Y}}||_2^2\\
    &\leq 2L^2 ||\mathbf{Y} - \hat{\mathbf{Y}}||_2^2 + \check{c}.
\end{align*}
Here, the second inequality comes from $\|\mathbf{a} + \mathbf{b}\|_2^2 \leq 2\|\mathbf{a}\|_2^2 + 2\|\mathbf{b}\|_2^2$, and $\check{c}$ is an upper bound of $2L^2 ||\hat{\mathbf{Y}} - \check{\mathbf{Y}}||_2^2$. This result indicates that minimizing the FR error improves image transmission quality.

\vspace{-3mm}
\section{Convergence Analysis}\label{sec:convergence analysis}
We first analyze how FR compensates for the compression error $\mathbf{m}_k^{\scriptstyle(t + 1)}$. To provide a comprehensive analysis, we formulate the update rules under three different scenarios for each iteration. Then, we delve into an in-depth investigation utilizing \textbf{Lemma \ref{lm:client sampling unbiasedness}} given below to analyze the convergence behavior of the proposed approach.

At global iteration $t$, the optimal global model achievable at the PS, assuming access to full, uncompressed local update information $\mathbf{g}_k^{\scriptstyle(t)}$ from all participating clients in $\mathcal{A}^{\scriptstyle(t)} = \mathcal{A}_m^{\scriptstyle(t)} \bigcup \mathcal{A}_o^{\scriptstyle(t)}$, is obtained by directly averaging these updates. This ideal model update is given by
\begin{align}\label{eq:best iteration}
    &\boldsymbol{w}^{\scriptstyle(t + 1)} = \boldsymbol{w}^{\scriptstyle(t)} - \frac{K}{K_m + K_o}\\
    &\quad\times\sum_{k\in\mathcal{A}^{\scriptstyle(t)}}p_k \left(\eta_c^{\scriptstyle(t)}\sum_{e = 0}^{E_c - 1}\nabla F_k^{\scriptstyle(t, e)}(\boldsymbol{w}_k^{\scriptstyle(t, e)}) + \mathbf{m}_k^{\scriptstyle(t)}\right).\nonumber
\end{align}  

Alternatively, when the PS does not utilize FR and instead relies solely on the compressed local updates $\bar{\mathbf{g}}_k^{\scriptstyle(t)}$ in $\mathcal{A}^{\scriptstyle(t)}$, the resulting global model update is expressed as
\begin{align}\label{eq:baseline iteration}
    &\boldsymbol{w}^{\scriptstyle(t + 1)} = \boldsymbol{w}^{\scriptstyle(t)} - \frac{K}{K_m + K_o}\\
    &\;\times\sum_{k\in\mathcal{A}^{\scriptstyle(t)}}p_k \left(\eta_c^{\scriptstyle(t)}\sum_{e = 0}^{E_c - 1}\nabla F_k^{\scriptstyle(t, e)}(\boldsymbol{w}_k^{\scriptstyle(t, e)}) + \mathbf{m}_k^{\scriptstyle(t)} - \mathbf{m}_k^{\scriptstyle(t + 1)}\right).\nonumber
\end{align}  

In contrast, the proposed FedSFR algorithm incorporates an additional model refinement step via FR at the PS, as outlined in \eqref{eq:server iteration}. This refinement enhances the global model beyond what is achievable with compressed gradients alone, which is verified in Section \ref{sec:experimental results}. The resulting update rule for FedSFR is
\begin{align}\label{eq:proposed iteration}
    &\boldsymbol{w}^{\scriptstyle(t + 1)} \!=\! \boldsymbol{w}^{\scriptstyle(t)} - \eta_s^{\scriptstyle(t)}\sum_{e = 0}^{E_s - 1}\!\nabla F_s^{\scriptstyle(t + \frac{1}{2}, e)}(\boldsymbol{w}_s^{\scriptstyle(t + \frac{1}{2}, e)}) - \frac{K}{K_m}\\
    &\;\times\sum_{k\in\mathcal{A}_m^{\scriptstyle(t)}}p_k \left(\eta_c^{\scriptstyle(t)}\sum_{e = 0}^{E_c - 1}\nabla F_k^{\scriptstyle(t, e)}(\boldsymbol{w}_k^{\scriptstyle(t, e)}) + \mathbf{m}_k^{\scriptstyle(t)} - \mathbf{m}_k^{\scriptstyle(t + 1)}\right).\nonumber
\end{align}

\begin{lemma}[\!\!\cite{li2019convergence}]\label{lm:client sampling unbiasedness}
Suppose we uniformly sample a subset $\mathcal{B}_0$ from a given set $\mathcal{B}$ without replacement. Then, the following unbiasedness property holds: $\mathbb{E}_{\mathcal{B}_0}\left[\frac{|\mathcal{B}|}{|\mathcal{B}_0|}\sum_{k\in\mathcal{B}_0}p_k x_k\right] = \sum_{k\in\mathcal{B}}p_k x_k,$ where $\sum_{k\in\mathcal{B}}p_k = 1$.
\begin{IEEEproof}
The proof builds upon Lemma $4$ in \cite{li2019convergence}. By transitioning the expectation computation from a subset-wise perspective to an element-wise perspective, we obtain
\begin{align*}
    \mathbb{E}_{\mathcal{B}_0}\left[\frac{|\mathcal{B}|}{|\mathcal{B}_0|}\sum_{k\in\mathcal{B}_0}p_k x_k\right] &= \frac{1}{\binom{|\mathcal{B}|}{|\mathcal{B}_0|}}\sum_{k\in\mathcal{B}}\binom{|\mathcal{B}| - 1}{|\mathcal{B}_0| - 1}\frac{|\mathcal{B}|}{|\mathcal{B}_0|}p_k x_k\\
    &= \sum_{k\in\mathcal{B}}p_k x_k.
\end{align*}
\end{IEEEproof}
\end{lemma}
\textbf{Lemma \ref{lm:client sampling unbiasedness}} demonstrates that the weighted summation of $x_k$ over a sampled subset $\mathcal{B}_0$, scaled by the factor $|\mathcal{B}|/|\mathcal{B}_0|$, has an expectation equal to the weighted summation of $x_k$ over the entire set $\mathcal{B}$. Leveraging this result, we analyze the expectations of the update rules in \eqref{eq:best iteration}, \eqref{eq:baseline iteration}, and \eqref{eq:proposed iteration} with respect to the sampling of $\mathcal{A}^{\scriptstyle(t)}$, i.e., $\mathbb{E}_{\mathcal{A}^{\scriptstyle(t)}}[\cdot]$. 

Applying \textbf{Lemma \ref{lm:client sampling unbiasedness}}, the difference between the optimal update in \eqref{eq:best iteration} and the baseline in \eqref{eq:baseline iteration} is expressed as $\sum_{k\in\mathcal{A}}p_k\mathbf{m}_k^{\scriptstyle(t + 1)}$, meaning the compression error. Similarly, the difference between \eqref{eq:best iteration} and our proposed update in \eqref{eq:proposed iteration} can be formulated as $\sum_{k\in\mathcal{A}}p_k\mathbf{m}_k^{\scriptstyle(t + 1)} - \eta_s^{\scriptstyle(t)}\sum_{e = 0}^{E_s - 1}\nabla F_s^{\scriptstyle(t + \frac{1}{2}, e)}(\boldsymbol{w}_s^{\scriptstyle(t + \frac{1}{2}, e)})$, where the additional term reflects the contribution of the server update process. This indicates that our proposed scheme is expected to better approximate the optimal global model in \eqref{eq:best iteration} compared to the baseline in \eqref{eq:baseline iteration}, which represents conventional FL incorporating compression and error feedback. To quantify the impact of the compression error and its relationship with the server update information, the following assumption is introduced.

\begin{prop}[Error memory compensation]\label{pro:error memory compensation}
For all $t$, let $\mathbf{a} = \sum_{k\in\mathcal{A}}p_k\mathbf{m}_k^{\scriptstyle(t + 1)}$ denote the compression error and $\mathbf{b} = \eta_s^{\scriptstyle(t)}\sum_{e = 0}^{E_s - 1}\nabla F_s^{\scriptstyle(t + \frac{1}{2}, e)}(\boldsymbol{w}_s^{\scriptstyle(t + \frac{1}{2}, e)})$ represent the server update information. Then, the inequality $\frac{\|\mathbf{a} - \mathbf{b}\|_2^2}{\|\mathbf{a}\|_2^2 + \|\mathbf{b}\|_2^2} \leq \varepsilon$ holds, where $\varepsilon$ is an upper bound such that $0 < \varepsilon \leq 2$.
\begin{IEEEproof}
    By the Cauchy–Schwarz inequality  such that $2(\|\mathbf{a}\|_2^2 + \|\mathbf{b}\|_2^2) = \sum_{i = 1}^D (1^2 + (-1)^2)(a_i^2 + b_i^2) \geq \sum_{i = 1}^D (1\cdot a_i + (-1)\cdot b_i)^2 = \|\mathbf{a} - \mathbf{b}\|_2^2$, we have $\|\mathbf{a} - \mathbf{b}\|_2^2 \leq 2(\|\mathbf{a}\|_2^2 + \|\mathbf{b}\|_2^2)$, which yields $\frac{\|\mathbf{a} - \mathbf{b}\|_2^2}{\|\mathbf{a}\|_2^2 + \|\mathbf{b}\|_2^2} \leq 2$. Thus, the upper bound $\varepsilon$ exists with $0 < \varepsilon \leq 2$. The value $\varepsilon = 2$ is achieved in the extreme case when $\mathbf{b} = - \mathbf{a}$.
\end{IEEEproof}
\end{prop}
If further assuming that $\varepsilon$ is smaller than $1$, \textbf{Proposition \ref{pro:error memory compensation}} directly leads to $\mathbf{a}^\textsf{T}\mathbf{b} \geq (1 - \varepsilon)\|\mathbf{a}\|_2\|\mathbf{b}\|_2$. Recall that $\mathbf{a}^\textsf{T}\mathbf{b} = \cos{(\varphi)}\|\mathbf{a}\|_2\|\mathbf{b}\|_2$, where $\varphi$ is an angle between $\mathbf{a}$ and $\mathbf{b}$. When $\varepsilon$ is sufficiently small, it follows that $\cos(\varphi) \approx 1$, indicating that $\mathbf{a}$ and $\mathbf{b}$ are nearly aligned in direction.  In this context, since $\mathbf{a}$ represents the compression error and $\mathbf{b}$ denotes the server update information, the update process at the server can effectively compensate for the compression error. Therefore, \textbf{Proposition \ref{pro:error memory compensation}} suggests that the impact of quantization or sparsification on update accuracy can be effectively modeled with $\varepsilon$.

Before deriving the convergence bound and rate of the proposed FedSFR method, we introduce several standard assumptions commonly adopted in the theoretical analysis of FL algorithms \cite{karimireddy2019error, li2019convergence}, as detailed below.

\begin{assume}\label{as:smoothness}
The local objective function $F_k(\boldsymbol{w})$ is $\beta_c$-smooth, and the global objective function $F(\boldsymbol{w})$, being the average of $\{F_k(\boldsymbol{w})\}_{k\in\mathcal{A}}$, is then also $\beta_c$-smooth. Furthermore, $F(\boldsymbol{w})$ is lower-bounded, i.e., $F(\boldsymbol{w}) \geq F(\boldsymbol{w}^*)$ for all $\boldsymbol{w}$.
\end{assume}

\begin{assume}\label{as:local gradient unbiasedness}
The stochastic gradient at each client is an unbiased estimator of the local gradient such as $\mathbb{E}[\nabla F_k^{\scriptstyle(t, e)}(\boldsymbol{w})] = \nabla F_k(\boldsymbol{w})$ for all $k, t, e$.
\end{assume}

\begin{assume}\label{as:gradient norm boundedness}
The expected squared norm of the stochastic gradient at each client and the PS is upper-bounded such as $\mathbb{E}[\|\nabla F_k^{\scriptstyle(t, e)}(\boldsymbol{w})\|_2^2] \leq G_k^2$ and $\mathbb{E}[\|\nabla F_s^{\scriptstyle(t, e)}(\boldsymbol{w})\|_2^2] \leq G_s^2$.
\end{assume}

\begin{thm}\label{thm:convergence rate}
Under \emph{\textbf{Proposition \ref{pro:error memory compensation}}}, \emph{\textbf{ Assumptions \ref{as:smoothness}, \ref{as:local gradient unbiasedness}, \ref{as:gradient norm boundedness}}}, and \emph{\textbf{Lemma \ref{lm:client sampling unbiasedness}}}, FedSFR with $\eta_c^{\scriptstyle(t)} = \alpha(t)/\sqrt{T}$ and $\eta_s^{\scriptstyle(t)} = \alpha(t)/T^{\frac{3}{4}}$, where $\alpha(t)$ is a monotonic decreasing function with an $\mathcal{O}(1)$ order, has an $\mathcal{O}(1/\sqrt{T})$ convergence rate:
\begin{align*}
    &\mathbb{E}\left[\frac{1}{T}\sum_{t = 0}^{T - 1}\|\nabla F(\boldsymbol{w}^{\scriptstyle(t)})\|_2^2\right] \leq \frac{1}{\sqrt{T}}\\
    &\quad\times \left(\frac{2}{\alpha(T)}(\mathbb{E}[F(\boldsymbol{w}^{\scriptstyle(0)})] - F(\boldsymbol{w}^*)) + \mathrm{A} + \frac{1}{\sqrt{T}}\mathrm{B} + \frac{1}{T}\mathrm{C}\right),
\end{align*}
where
\begin{align*}
    \mathrm{A} &= 2\alpha(0)\beta_c\frac{K}{K_m}E_c^2 G_{k, max}^2 + \frac{\alpha(0)^2}{\alpha(T)^2}\frac{E_s^2}{E_c}G_s^2,\\
    \mathrm{B} &= \alpha(0)^2\beta_c^2\left\{\left(\frac{(K - K_m)^2}{K_m^2} + \varepsilon\right)\frac{16(1 - \nu)}{\nu^2} + \frac{2}{3}\right\}\\
    &\qquad\times E_c^3 G_{k, max}^2 + 2\frac{\alpha(0)^2}{\alpha(T)}\beta_c E_s^2 G_s^2,\\
    \mathrm{C} &= 4\alpha(0)^2\varepsilon\beta_c^2 E_c E_s^2 G_s^2,
\end{align*}
$G_{k, max}^2 = \max_k G_k^2$, and $0 < \nu \leq 1$.
\begin{IEEEproof}
    See \emph{Appendix \ref{apdx:convergence rate}}.
\end{IEEEproof}
\end{thm}

According to \textbf{Theorem \ref{thm:convergence rate}}, as the number of global iterations $T$ increases, the right-hand side of the derived bound asymptotically approaches zero, under the condition that $\eta_s^{\scriptstyle(t)} < \eta_c^{\scriptstyle(t)}$.  A smaller server learning rate is reasonable, since the main objective is accurate image reconstruction rather than exact feature reconstruction. Consequently, the global model at iteration $T - 1$, denoted by $\boldsymbol{w}^{\scriptstyle(T)}$, converges with respect to the global objective function $F(\boldsymbol{w})$. This convergence directly relates to the reconstruction accuracy of the JSCC encoder-decoder pair and the VQ codebook. Furthermore, the convergence bound highlights the role of $\varepsilon$: a smaller $\varepsilon$ leads to faster convergence. In particular, when the server-side update process effectively compensates for the compression error, i.e., when \textbf{Proposition \ref{pro:error memory compensation}} is well formulated, the convergence of the global model is significantly accelerated.

\vspace{-3mm}
\section{Experimental Results}\label{sec:experimental results}
The performance of FedSFR is evaluated on the CIFAR-10 dataset, using the peak signal-to-noise ratio (PSNR) as the evaluation metric. We set $K = 50$, $|\mathcal{D}_k| = 800$, $K_m = 10$, $K_o = 10$, $S_m/D = 0.2$, and $S_o/D = 0.1$. The mini-batch size is set to $16$, and both the clients and the PS perform 3 epochs of training per global iteration, with the total number of global iterations set to $T = 100$. The learning rates are initialized as $\eta_c^{\scriptstyle(0)} = 0.01$ for the clients and $\eta_s^{\scriptstyle(0)} = 0.0001$ for the PS, both of which are decayed by a factor of 0.9 every 10 global iterations. The NN architecture of the JSCC encoder and decoder is adapted from \cite{huh2025universal} with a margin parameter of $\mu = 0.1$, and consists of approximately $0.51$M parameters. The number of feature vectors is set to $N = 256$, and 256-QAM is employed. Accordingly, each client in $\mathcal{A}_o^{\scriptstyle(t)}$ transmits $100$($=|\mathcal{P}_k|$) instances of feature vectors in $\mathcal{Y}_k^{\scriptstyle(t)}$ to the PS at each global iteration. Unless otherwise specified, the semantic communication modules are trained under an AWGN channel for semantic transmission with 20dB SNR and $\alpha = 0.05$, while the PSNR is evaluated at the same SNR level.

We compare FedSFR with five baselines. The first three differ in their design of aggregation weights, while the remaining two are based on KD techniques:
\begin{itemize}
\item FedAvg \cite{mcmahan2017communication}: This method uses the $\mathsf{Compress}$ function in \eqref{eq:compression} along with an error feedback mechanism for the participants in $\mathcal{A}^{\scriptstyle(t)}$, without applying the proposed server-side update process. All other settings are identical to FedSFR, except for those related to the PS.
\item FedDMA \cite{xing2023multi} and FedLol \cite{nguyen2024efficient}: These approaches differ from FedAvg only in the design of aggregation weights, which are client-dependent and non-uniform.
\item SKBS \cite{lu2024efficient}: This method performs KD between the local model and the aggregated server model, where the latter is smaller. We configure the server model to have approximately $0.13$M parameters. To match this setting, we adjust the sparsification levels to $S_m/D = 0.8$ and $S_o/D = 0.4$.
\item FedSFD \cite{xu2024federated}: This method employs semantic feature distillation for all the participants in $\mathcal{A}^{\scriptstyle(t)}$, using the personalized datasets $\{\mathcal{P}_k\}_{k\in\mathcal{A}}$ and transmitting 50 pairs of encoder output features and intermediate decoder features. Notably, the uplink communication cost is significantly higher than that of other methods. The PS utilizes a larger JSCC decoder model than the clients.
\end{itemize}

\begin{figure}[!t]
    \centering
    \setlength{\subfigcapskip}{-3mm}
    \subfigure[Comparison with baselines]{\includegraphics[width=0.7\linewidth]{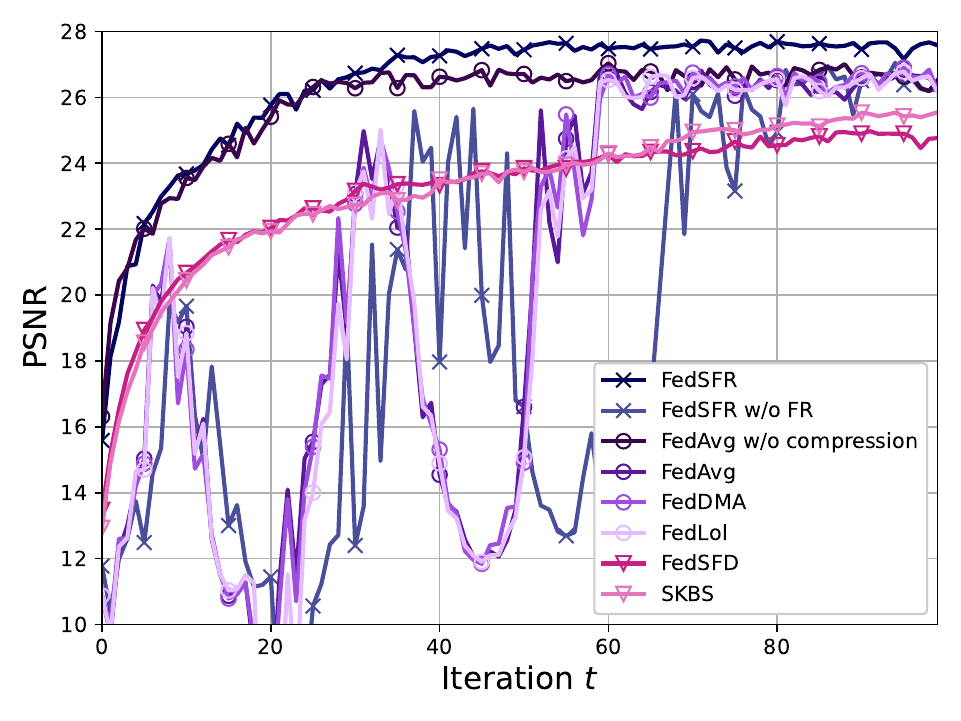}
    \label{fig:stability cifar}}\vspace{-2mm}
    \subfigure[Learning rates]{\includegraphics[width=0.7\linewidth]{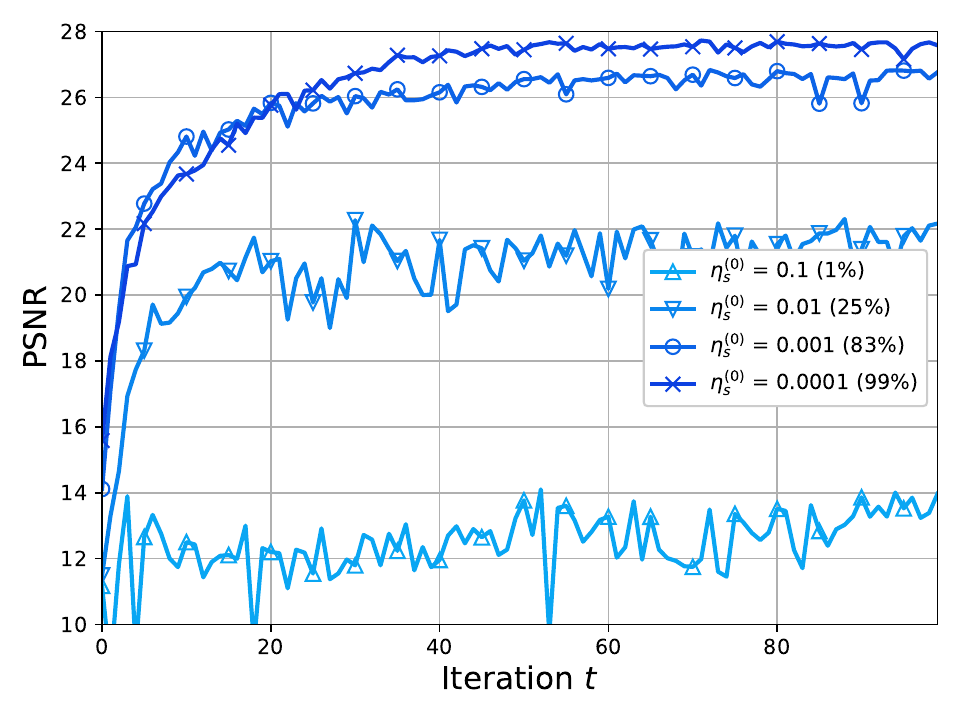}\label{fig:learning rate cifar}}\vspace{-2mm}
    \caption{PSNR of (a) the proposed scheme and the baselines, and (b) the proposed scheme with varying learning rates for CIFAR-10 dataset.}
    \vspace{-5mm}
    \label{fig:cifar}
\end{figure}

\begin{figure}[!t]
    \centering
    \includegraphics[width=0.7\linewidth]{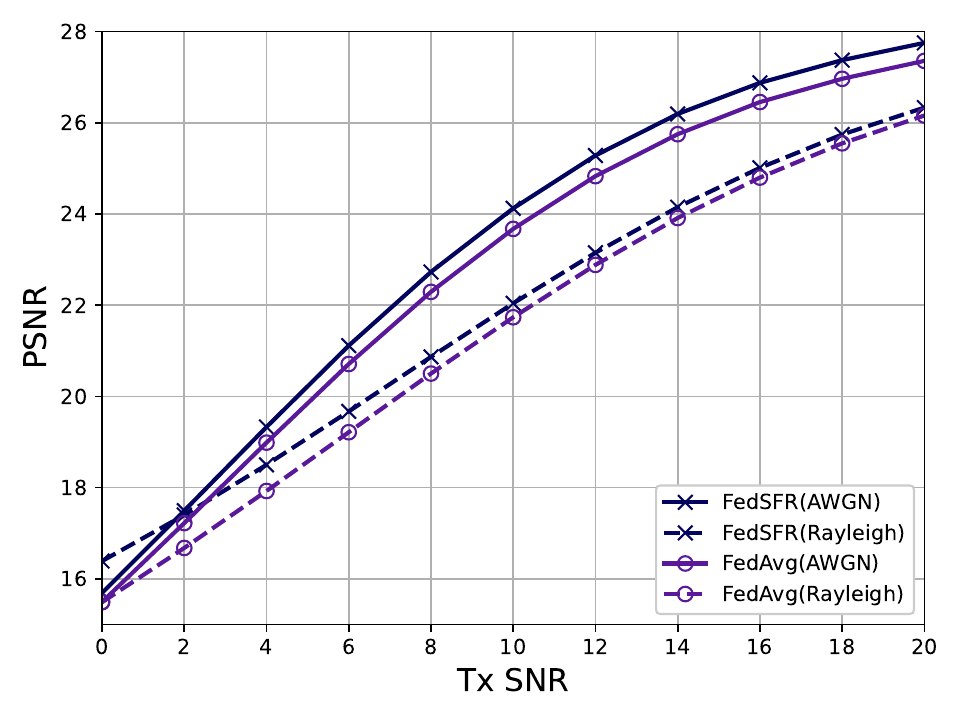}
    \vspace{-4mm}
    \caption{PSNR of JSCC models trained via FedSFR and FedAvg, tested over AWGN and Rayleigh fading channels for CIFAR-10 dataset.}
    \vspace{-5mm}
    \label{fig:inference}
\end{figure}

\textbf{Comparison with Baselines.}\quad Fig. \ref{fig:stability cifar} presents the PSNR performance curves of FedSFR and baseline methods. The three aggregation weight design approaches show noticeable instability, primarily due to the use of sparsified local updates when aggregating the global model $\boldsymbol{w}^{\scriptstyle(t)}$. This is because, unlike classification tasks that rely on categorical outputs derived from softmax-based distributions, regression tasks such as image reconstruction demand higher fidelity in the output. The two KD-aided approaches, on the other hand, yield the lowest PSNR and exhibit the slowest convergence, indicating that the KD strategy is insufficiently effective in FL for image semantic communication. In contrast, FedSFR demonstrates consistent and stable training behavior. This is attributed to the FR step performed at the PS, which further enhances the quality of the aggregated model. This refinement process successfully reduces the compression-induced error $\mathbf{m}_k^{\scriptstyle(t + 1)}$, aligning with \textbf{Proposition \ref{pro:error memory compensation}}. This positive impact is also captured in the parameter $\varepsilon$ in \textbf{Theorem \ref{thm:convergence rate}}. Thanks to this enhanced stability, FedSFR reaches a higher level of reconstruction quality more rapidly than baseline methods, especially during the initial phase of training. Surprisingly, FedSFR even surpasses the converged performance of FedAvg without compression. However, as the FedAvg update relies only on clients in $\mathcal{A}_m^{\scriptstyle(t)}$, FedSFR without FR exhibits unstable training and slower convergence, highlighting the importance of the FR refinement step.

\textbf{Learning Rates.}\quad Fig. \ref{fig:learning rate cifar} shows the effect of different initial server learning rates $\eta_s^{\scriptstyle(0)}$, selected from the set $\{0.1, 0.01, 0.001, 0.0001\}$. To observe whether the FR step at the PS leads to performance gains, we measure the improvement ratio, indicated in the legend, which quantifies the fraction of successful global iterations specifically by comparing the model before and after FR, i.e., $\boldsymbol{w}_s^{\scriptstyle(t + \frac{1}{2}, 0)}$ versus $\boldsymbol{w}_s^{\scriptstyle(t + \frac{1}{2}, E_s)}$. According to the condition established in \textbf{Theorem \ref{thm:convergence rate}}, $\eta_s^{\scriptstyle(t)}$ should remain smaller than $\eta_c^{\scriptstyle(t)}$ to ensure convergence. Larger server learning rates, particularly $\eta_s^{\scriptstyle(0)} = 0.1$ and $0.01$, result in degraded performance, because the server-side FR dominates the learning dynamics, impeding effective image reconstruction. In contrast, smaller values such as $\eta_s^{\scriptstyle(0)} = 0.001$ and $0.0001$ yield more stable training, a higher frequency of beneficial FR steps, and improved final performance. These observations highlight the importance of properly selecting the learning rates to satisfy $\eta_s^{\scriptstyle(t)} < \eta_c^{\scriptstyle(t)}$, as suggested by \textbf{Theorem \ref{thm:convergence rate}}, to maximize the benefit of FR in FedSFR.

\textbf{Performance of the trained modules.}\quad Fig. \ref{fig:inference} evaluates the PSNR of the trained semantic communication modules under two different semantic transmission environments, i.e., AWGN and Rayleigh fading channels. Specifically, for fading scenarios, we divide the transmitted $N = 256$ feature vectors into groups of size $64$, where each group experiences an independently sampled Rayleigh fading channel with gain $h\sim\mathcal{CN}(0, 1)$. The same channel setting is applied during both training and inference. The transmit SNR, i.e., the x-axis in Fig. \ref{fig:inference}, is fixed at $\frac{P}{\sigma^2}$, while the received SNR varies according to $\frac{|h|^2P}{\sigma^2}$. The reported PSNR is averaged over $1,000$ independently sampled Rayleigh fading channel realizations for each transmit SNR level. Fig. \ref{fig:inference} demonstrates that, under AWGN channels, the JSCC models trained via FedSFR consistently outperform those trained with baseline FL methods, reflecting the advantages already observed during training. Moreover, under Rayleigh fading channels, FedSFR continues to achieve higher PSNR across all SNR levels, confirming that the performance gains provided by FedSFR are preserved even under channel impairments.

\begin{figure}[!t]
    \centering
    \setlength{\subfigcapskip}{-3mm}
    \subfigure[Comparison with baselines]{\includegraphics[width=0.7\linewidth]{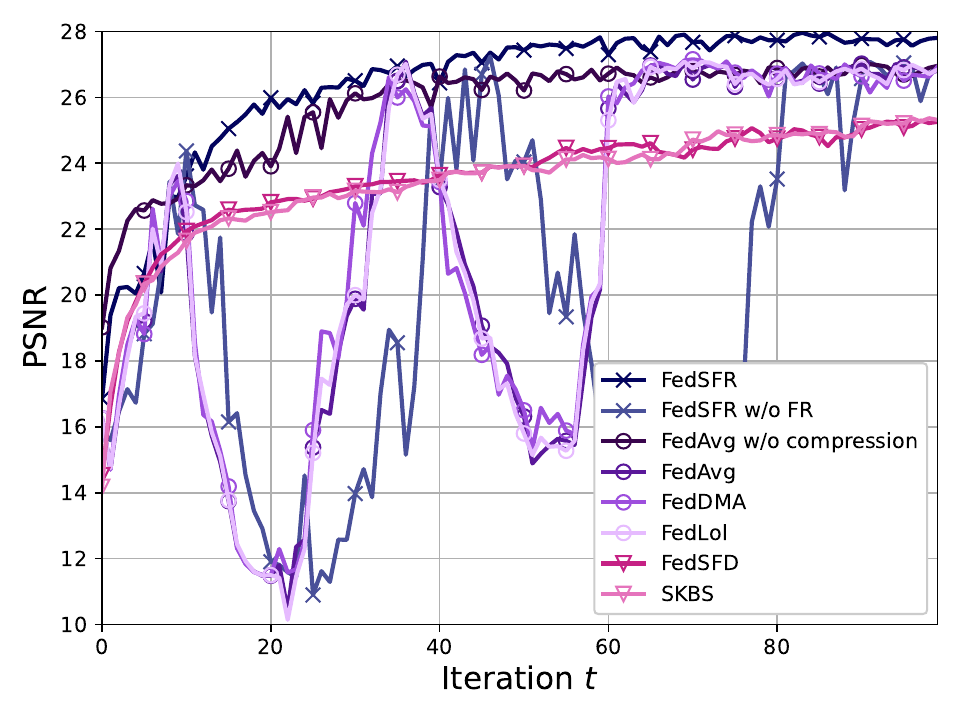}
    \label{fig:stability celeb}}\vspace{-2mm}
    \subfigure[Learning rates]{\includegraphics[width=0.7\linewidth]{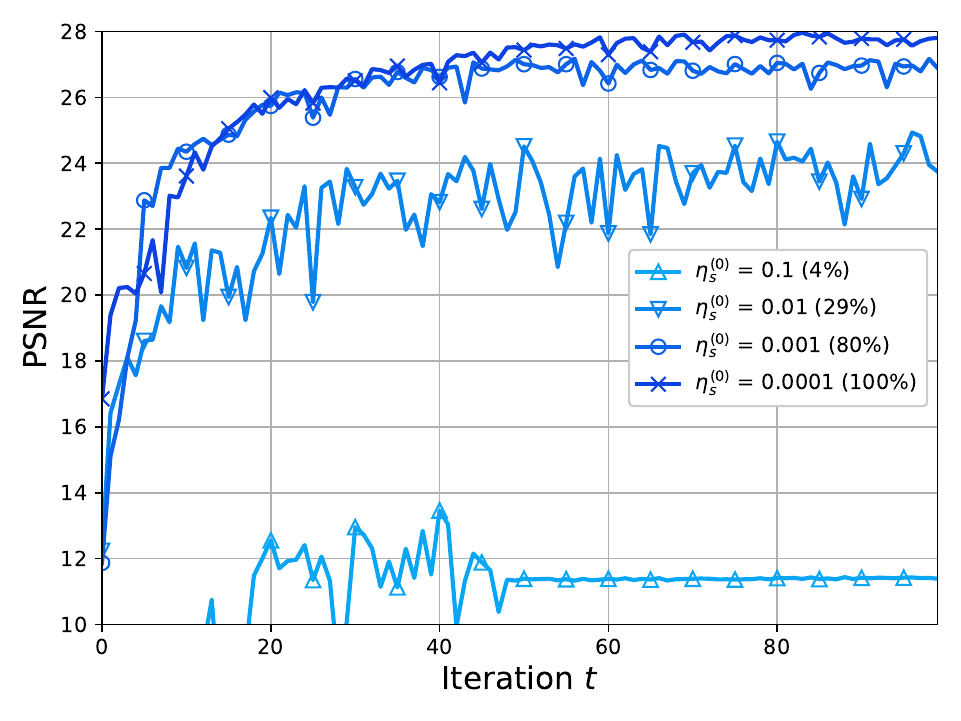}\label{fig:learning rate celeb}}\vspace{-2mm}
    \caption{PSNR of (a) the proposed scheme and the baselines, and (b) the proposed scheme with varying learning rates for CelebA dataset.}
    \vspace{-5mm}
    \label{fig:celeb}
\end{figure}

\textbf{Evaluation with Higher Resolution Images.}\quad To verify the benefit of FedSFR for higher resolution images, as depicted in Fig. \ref{fig:celeb}, we replicate all previous experiments using the CelebA dataset. The number of model parameters is increased to about $0.95$M with $N = 1024$, leading to transmitting $50$ instances of feature vectors per global iteration in FedSFR. Accordingly, the server model size of SKBS is varied to about $0.24$M, thereby setting $S_m/D = 0.78$ and $S_o/D = 0.39$, while FedSFD uploads $10$ pairs of features. The results in Fig. \ref{fig:celeb} follow a pattern consistent with those obtained using the CIFAR-10 dataset, as illustrated in Fig. \ref{fig:cifar}, underscoring the robustness and consistency of FedSFR across different datasets and image resolutions.

\begin{figure}[!t]
    \centering
    \setlength{\subfigcapskip}{-3mm}
    \subfigure[$10$dB training SNR]{\includegraphics[width=0.7\linewidth]{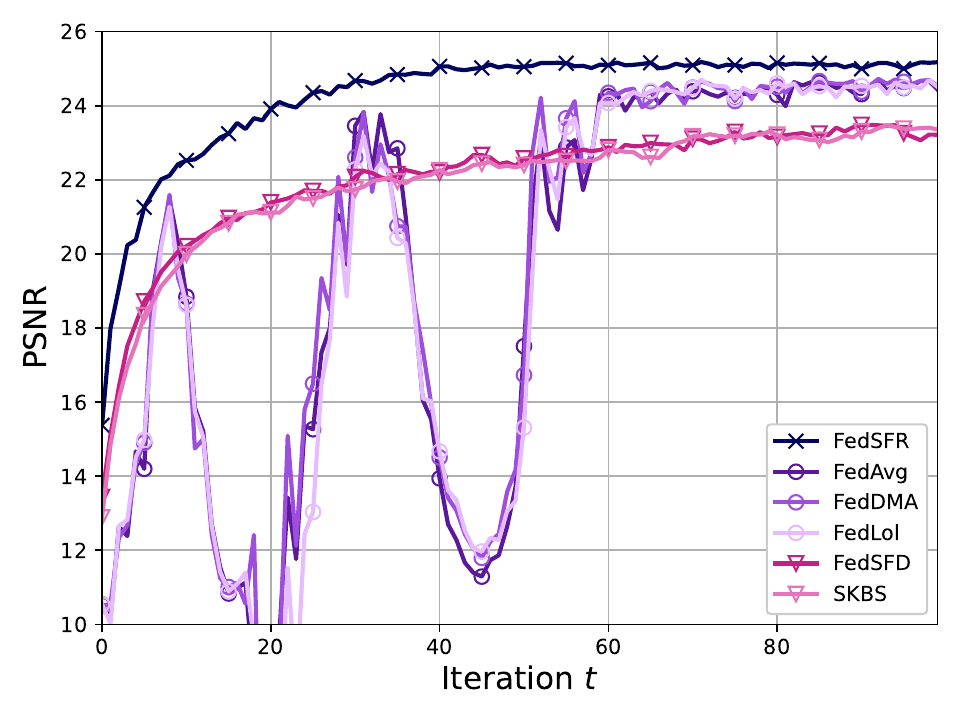}\label{fig:10db cifar}}\vspace{-2mm}
    \subfigure[Non-IID]{\includegraphics[width=0.7\linewidth]{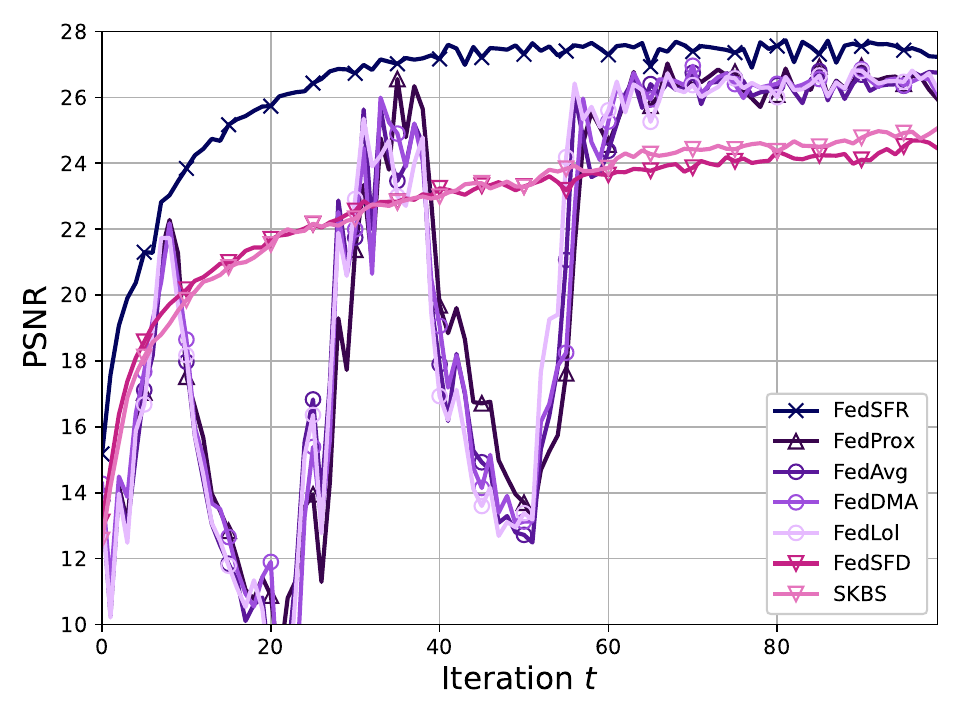}
    \label{fig:non-iid cifar}}\vspace{-2mm}
    \subfigure[Strategy selection]{\includegraphics[width=0.7\linewidth]{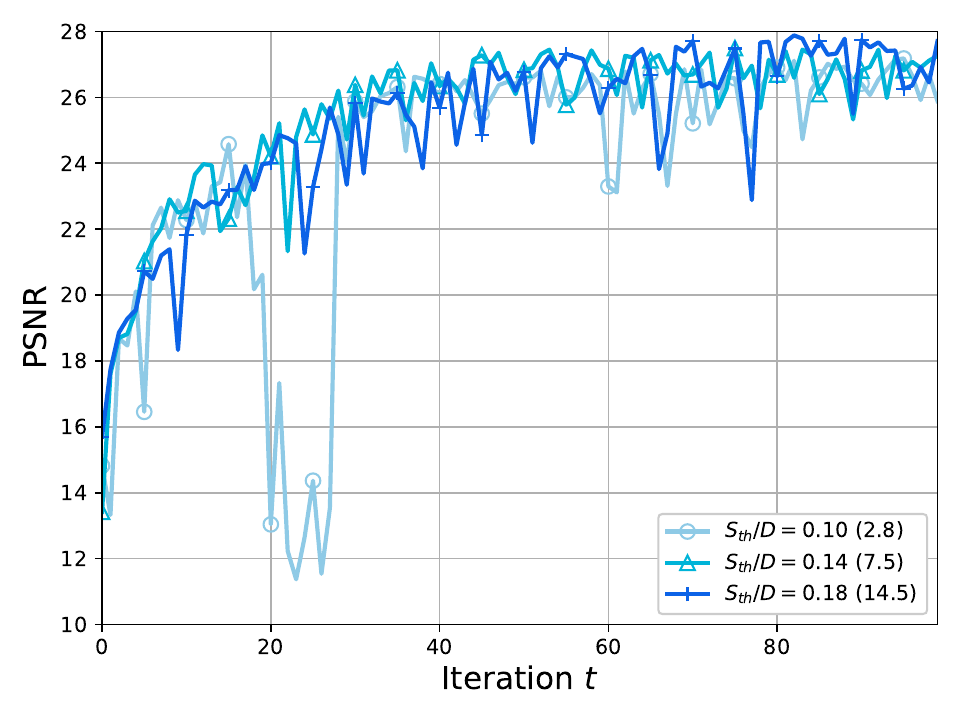}\label{fig:threshold cifar}}\vspace{-2mm}
    \caption{PSNR of the proposed and baseline schemes (a) under $10$dB training SNR, (b) under non-IID data, and (c) the proposed scheme with varying strategy selection thresholds for CIFAR-10 dataset.}
    \vspace{-5mm}
    \label{fig:ablation}
\end{figure}

\textbf{Ablation Studies.}\quad To evaluate the robustness of FedSFR under different semantic communication environments, we first evaluate its performance with a lower training SNR of $10$dB and non-independent and identically distributed (IID) local/public data distributions. Note that a Dirichlet distribution with parameter $0.5$ is used to implement the non-IID setting. Figs. \ref{fig:10db cifar} and \ref{fig:non-iid cifar} show that FedSFR consistently maintains superior PSNR and stable training behavior compared to the baselines in line with the trends observed at $20$dB SNR with IID data in Fig. \ref{fig:stability cifar}. In particular, to mitigate data heterogeneity, FedProx \cite{li2020federated} with a proximal regularization term weighted by $\mu=0.01$ is incorporated as a baseline in Fig. \ref{fig:non-iid cifar}. These results confirm the robustness of FedSFR to both channel noise and data heterogeneity. Moreover, as the PSNR values under the non-IID setting are comparable to those under the IID setting, it can be inferred that feature vectors generated from the public dataset effectively convey characteristics of the private data distributions of clients in $\mathcal{A}_o^{\scriptstyle(t)}$.

We further investigate the impact of the strategy selection threshold $S_\mathsf{th}$ under dynamic wireless channel conditions. Specifically, the number of available uplink bits $B_k$ is varied at each iteration in proportion to the channel capacity between the clients and the PS, where the channel SNR follows an exponential distribution with scale parameter 1. The averaged size of $|\mathcal{A}_o^{\scriptstyle(t)}|$ over $T$ iterations is reported in the legend. As shown in Fig. \ref{fig:threshold cifar}, when $S_\mathsf{th}/D = 0.10$, training is initially unstable due to weaker FR effects from fewer clients in $\mathcal{A}_o^{\scriptstyle(t)}$. For $S_\mathsf{th}/D = 0.18$, peak PSNR is highest, but performance becomes unstable later because excessive FR occurs with fewer clients in $\mathcal{A}_m^{\scriptstyle(t)}$. In contrast, the setting $S_\mathsf{th}/D = 0.14$ achieves a more balanced training process, likely because the numbers of clients in $\mathcal{A}_o^{\scriptstyle(t)}$ and $\mathcal{A}_m^{\scriptstyle(t)}$ remain relatively balanced throughout training, resulting in stable and sufficiently high PSNR. These observations imply that maintaining balanced group sizes is beneficial for stable training performance, motivating adaptive threshold control under time-varying wireless environments as a promising direction for future work.

\vspace{-3mm}
\section{Conclusion}\label{sec:conclusion}
We introduced the FedSFR algorithm, a digital communication framework for FL that incorporates FR learning and leverages the shared fixed VQ codebook for both the semantic communication and the JSCC model update by FedSFR, enabling efficient image transmission. By incorporating an additional server-side update process based on FR, FedSFR achieves both effective model training and efficient information transmission. To facilitate effective FR learning, we proposed a novel FR procedure tailored to VQ-based systems, which is intuitively motivated by and analogous to the image reconstruction process. Moreover, we rigorously derived the convergence rate of the proposed algorithm, detailing its design and providing theoretical guarantees. Through extensive simulations, we demonstrated that the FedSFR framework outperforms existing algorithms in terms of convergence rate, task performance, training stability, and communication efficiency. By mitigating the impact of compression errors through additional server-side updates, our approach offers a scalable and practical solution for real-world digital image semantic communication applications.

\vspace{-3mm}
\appendices
\section{Proof of Theorem \ref{thm:convergence rate}}\label{apdx:convergence rate}
Let $\tilde{\boldsymbol{w}}^{\scriptstyle(t)}$ be an auxiliary virtual sequence, such as
\begin{align} \label{eq:virtual_seq}
    \tilde{\boldsymbol{w}}^{\scriptstyle(t + 1)} = \tilde{\boldsymbol{w}}^{\scriptstyle(t)} &- \frac{K}{K_m}\sum_{k\in\mathcal{A}_m^{\scriptstyle(t)}}p_k \eta_c^{\scriptstyle(t)}\sum_{e = 0}^{E_c - 1}\nabla F_k^{\scriptstyle(t, e)}(\boldsymbol{w}_k^{\scriptstyle(t, e)})\nonumber\\
    &- \eta_s^{\scriptstyle(t - 1)}\sum_{e = 0}^{E_s - 1}\nabla F_s^{\scriptstyle(t - \frac{1}{2}, e)}(\boldsymbol{w}_s^{\scriptstyle(t - \frac{1}{2}, e)}),
\end{align}
where we define that $\nabla F_s^{\scriptstyle(t - \frac{1}{2}, e)}(\boldsymbol{w}_s^{\scriptstyle(t - \frac{1}{2}, e)}) = \mathbf{0}$ for $t = 0$ and all $e$, and $\tilde{\boldsymbol{w}}^{\scriptstyle(0)} = \boldsymbol{w}^{\scriptstyle(0)}$ by definition. Then, the relation between $\tilde{\boldsymbol{w}}^{\scriptstyle(t)}$ and $\boldsymbol{w}^{\scriptstyle(t)}$ can be given as
\begin{align*}
    &\tilde{\boldsymbol{w}}^{\scriptstyle(t)} = \boldsymbol{w}^{\scriptstyle(t)}\\
    &\; - \left[\frac{K}{K_m}\sum_{k\in\mathcal{A}}p_k\mathbf{m}_k^{\scriptstyle(t)} - \eta_s^{\scriptstyle(t - 1)}\sum_{e = 0}^{E_s - 1}\nabla F_s^{\scriptstyle(t - \frac{1}{2}, e)}(\boldsymbol{w}_s^{\scriptstyle(t - \frac{1}{2}, e)})\right].
\end{align*}
Under \textbf{Assumption \ref{as:smoothness}}, 
\begin{align}\label{eq:initial inequality}
    \mathbb{E}[F(\tilde{\boldsymbol{w}}^{\scriptstyle(t + 1)}) - F(\tilde{\boldsymbol{w}}^{\scriptstyle(t)})] &\leq \underbrace{\mathbb{E}[\nabla F(\tilde{\boldsymbol{w}}^{\scriptstyle(t)})^\textsf{T}(\tilde{\boldsymbol{w}}^{\scriptstyle(t + 1)} - \tilde{\boldsymbol{w}}^{\scriptstyle(t)})]}_{(a)}\nonumber\\
    &\; + \frac{\beta_c}{2}\underbrace{\mathbb{E}[\|\tilde{\boldsymbol{w}}^{\scriptstyle(t + 1)} - \tilde{\boldsymbol{w}}^{\scriptstyle(t)}\|_2^2]}_{(b)}.
\end{align}

Now we can derive an upper bound of $(a)$ in \eqref{eq:initial inequality} as follows:
\begin{align*} 
    (a) &=-\eta_c^{\scriptstyle(t)}\sum_{e = 0}^{E_c - 1}\mathbb{E}\left[\nabla F(\tilde{\boldsymbol{w}}^{\scriptstyle(t)})^\textsf{T}\sum_{k\in\mathcal{A}}p_k\nabla F_k(\boldsymbol{w}_k^{\scriptstyle(t, e)})\right]\\
    &\quad - \eta_s^{\scriptstyle(t - 1)}\sum_{e = 0}^{E_s - 1}\mathbb{E}[\nabla F(\tilde{\boldsymbol{w}}^{\scriptstyle(t)})^\textsf{T}\nabla F_s^{\scriptstyle(t - \frac{1}{2}, e)}(\boldsymbol{w}_s^{\scriptstyle(t - \frac{1}{2}, e)})]\\
    &\leq -\frac{\eta_c^{\scriptstyle(t)}}{2}E_c\mathbb{E}[\|\nabla F(\tilde{\boldsymbol{w}}^{\scriptstyle(t)})\|_2^2]\\
    &\quad - \frac{\eta_c^{\scriptstyle(t)}}{2}\sum_{e = 0}^{E_c - 1}\mathbb{E}\left[\left\|\sum_{k\in\mathcal{A}}p_k\nabla F_k(\boldsymbol{w}_k^{\scriptstyle(t, e)})\right\|_2^2\right]\\
    &\quad + \frac{\eta_c^{\scriptstyle(t)}}{2}\sum_{e = 0}^{E_c - 1}\mathbb{E}\left[\left\|\nabla F(\tilde{\boldsymbol{w}}^{\scriptstyle(t)}) - \sum_{k\in\mathcal{A}}p_k\nabla F_k(\boldsymbol{w}_k^{\scriptstyle(t, e)})\right\|_2^2\right]\\
    &\quad + \frac{\eta_s^{\scriptstyle(t - 1)}}{2}E_s\kappa\mathbb{E}[\|\nabla F(\tilde{\boldsymbol{w}}^{\scriptstyle(t)})\|_2^2]\\
    &\quad + \frac{\eta_s^{\scriptstyle(t - 1)}}{2}\sum_{e = 0}^{E_s - 1}\frac{1}{\kappa}\mathbb{E}[\|\nabla F_s^{\scriptstyle(t - \frac{1}{2}, e)}(\boldsymbol{w}_s^{\scriptstyle(t - \frac{1}{2}, e)})\|_2^2]\\
    &\leq -\frac{\eta_c^{\scriptstyle(t)}}{2}\underbrace{\sum_{e = 0}^{E_c - 1}\mathbb{E}\left[\left\|\sum_{k\in\mathcal{A}}p_k\nabla F_k(\boldsymbol{w}_k^{\scriptstyle(t, e)})\right\|_2^2\right]}_{(a.1)}\\
    &\quad + \frac{\eta_c^{\scriptstyle(t)}}{2}\underbrace{\sum_{e = 0}^{E_c - 1}\mathbb{E}\left[\left\|\nabla F(\tilde{\boldsymbol{w}}^{\scriptstyle(t)}) - \sum_{k\in\mathcal{A}}p_k\nabla F_k(\boldsymbol{w}_k^{\scriptstyle(t, e)})\right\|_2^2\right]}_{(a.2)}\\
    &\quad + \frac{(\eta_s^{\scriptstyle(t - 1)})^2 E_s^2}{2\eta_c^{\scriptstyle(t)} E_c}G_s^2,
\end{align*}
where the first equality comes from \textbf{Lemma \ref{lm:client sampling unbiasedness}} and \textbf{Assumption \ref{as:local gradient unbiasedness}}; the first inequality is based on two simple inequalities, such as $-\mathbf{a}^\textsf{T}\mathbf{b} = \frac{1}{2}(-\|\mathbf{a}\|_2^2 - \|\mathbf{b}\|_2^2 + \|\mathbf{a - b}\|_2^2)$ and $-\mathbf{a}^\textsf{T}\mathbf{b} \leq \frac{1}{2}(\kappa\|\mathbf{a}\|_2^2 + \frac{1}{\kappa}\|\mathbf{b}\|_2^2)$; the last inequality comes from \textbf{Assumption \ref{as:gradient norm boundedness}} while setting $\kappa = \frac{\eta_c^{\scriptstyle(t)}}{\eta_s^{\scriptstyle(t - 1)}}\cdot\frac{E_c}{E_s}$.  

Next, we  get a lower bound of $(a.1)$ by choose only $e = 0$ from the summation of positive norm values:
\begin{align*}
    (a.1) &\geq \mathbb{E}\left[\left\|\sum_{k\in\mathcal{A}}p_k\nabla F_k(\boldsymbol{w}_k^{\scriptstyle(t, 0)})\right\|_2^2\right]\\
    &= \mathbb{E}\left[\left\|\sum_{k\in\mathcal{A}}p_k\nabla F_k(\boldsymbol{w}^{\scriptstyle(t)})\right\|_2^2\right] = \mathbb{E}[\|\nabla F(\boldsymbol{w}^{\scriptstyle(t)})\|_2^2].
\end{align*} 

We use \textbf{Lemma \ref{lm:upper bound of (a.2)}} to get an upper bound of $(a.2)$.
\begin{lemma}\label{lm:upper bound of (a.2)}
Under \emph{\textbf{Assumptions \ref{as:smoothness} and \ref{as:gradient norm boundedness}}}, the inequality below is satisfied:
\begin{align*}
    &\sum_{e = 0}^{E_c - 1}\mathbb{E}\left[\left\|\nabla F(\tilde{\boldsymbol{w}}^{\scriptstyle(t)}) - \sum_{k\in\mathcal{A}}p_k\nabla F_k(\boldsymbol{w}_k^{\scriptstyle(t, e)})\right\|_2^2\right]\leq 4E_c\beta_c^2\\
    &\quad\times \Biggl\{\left(\frac{(K - K_m)^2}{K_m^2} + \varepsilon\right)\frac{4(1 - \nu)}{\nu^2}(\eta_c^{\scriptstyle(0)})^2 E_c^2\sum_{k\in\mathcal{A}}p_k G_k^2\\
    &\quad\qquad + \varepsilon(\eta_s^{\scriptstyle(t - 1)})^2 E_s^2 G_s^2\Biggl\} + \frac{2}{3}E_c^3\beta_c^2(\eta_c^{\scriptstyle(t)})^2\sum_{k\in\mathcal{A}}p_k G_k^2.
\end{align*}
\begin{IEEEproof}
    See \emph{Appendix \ref{apdx:upper bound of (a.2)}}.
\end{IEEEproof}
\end{lemma}

Finally, using \eqref{eq:virtual_seq} and a simple inequality such as $\|\mathbf{a} + \mathbf{b}\|_2^2 \leq 2(\|\mathbf{a}\|_2^2 + \|\mathbf{b}\|_2^2)$, we get an upper bound of $(b)$ in \eqref{eq:initial inequality}:
\begin{align*}
    &(b)\leq 2\mathbb{E}\left[\left\|\frac{K}{K_m}\sum_{k\in\mathcal{A}_m^{\scriptstyle(t)}}p_k \eta_c^{\scriptstyle(t)}\sum_{e = 0}^{E_c - 1}\nabla F_k^{\scriptstyle(t, e)}(\boldsymbol{w}_k^{\scriptstyle(t, e)})\right\|_2^2\right]\\
    &\qquad + 2\mathbb{E}\left[\left\|\eta_s^{\scriptstyle(t - 1)}\sum_{e = 0}^{E_s - 1}\nabla F_s^{\scriptstyle(t - \frac{1}{2}, e)}(\boldsymbol{w}_s^{\scriptstyle(t - \frac{1}{2}, e)})\right\|_2^2\right]\\
    &\quad\leq 2\frac{K}{K_m}(\eta_c^{\scriptstyle(t)})^2E_c\\
    &\quad\qquad\times \mathbb{E}\left[\frac{K}{K_m}\sum_{k\in\mathcal{A}_m^{\scriptstyle(t)}}p_k\sum_{e = 0}^{E_c - 1}\left\|\nabla F_k^{\scriptstyle(t, e)}(\boldsymbol{w}_k^{\scriptstyle(t, e)})\right\|_2^2\right]\\
    &\qquad + 2(\eta_s^{\scriptstyle(t - 1)})^2 E_s\mathbb{E}\left[\sum_{e = 0}^{E_s - 1}\left\|\nabla F_s^{\scriptstyle(t - \frac{1}{2}, e)}(\boldsymbol{w}_s^{\scriptstyle(t - \frac{1}{2}, e)})\right\|_2^2\right]\\
    &\quad\leq 2\frac{K}{K_m}(\eta_c^{\scriptstyle(t)})^2 E_c^2\sum_{k\in\mathcal{A}}p_k G_k^2 + 2(\eta_s^{\scriptstyle(t - 1)})^2 E_s^2 G_s^2,
\end{align*}  
where the second inequality is due to Jensen's inequality and the last inequality comes from \textbf{Assumption \ref{as:gradient norm boundedness}} and \textbf{Lemma \ref{lm:client sampling unbiasedness}}.

Substituting the above inequalities, which contain various bounds, to (\ref{eq:initial inequality}), we can derive
\begin{align*}
    &\mathbb{E}[F(\tilde{\boldsymbol{w}}^{\scriptstyle(t + 1)}) - F(\tilde{\boldsymbol{w}}^{\scriptstyle(t)})] \leq -\frac{\eta_c^{\scriptstyle(t)}}{2}\mathbb{E}[\|\nabla F(\boldsymbol{w}^{\scriptstyle(t)})\|_2^2]\\
    &\;\; + \frac{\eta_c^{\scriptstyle(t)}}{2}\cdot 4E_c\beta_c^2\\
    &\qquad\times \Biggl\{\left(\frac{(K - K_m)^2}{K_m^2} + \varepsilon\right)\frac{4(1 - \nu)}{\nu^2}(\eta_c^{\scriptstyle(0)})^2 E_c^2G_{k, max}^2\\
    &\qquad\qquad + \varepsilon(\eta_s^{\scriptstyle(t - 1)})^2 E_s^2 G_s^2\Biggl\} + \frac{\eta_c^{\scriptstyle(t)}}{2}\cdot\frac{2}{3}E_c^3\beta_c^2(\eta_c^{\scriptstyle(t)})^2 G_{k, max}^2\\
    &\;\; + \frac{(\eta_s^{\scriptstyle(t - 1)})^2 E_s^2}{2\eta_c^{\scriptstyle(t)} E_c}G_s^2\\
    &\;\; + \frac{\beta_c}{2}\left(2\frac{K}{K_m}(\eta_c^{\scriptstyle(t)})^2 E_c^2 G_{k, max}^2 + 2(\eta_s^{\scriptstyle(t - 1)})^2 E_s^2 G_s^2\right),
\end{align*}
where $G_{k, max}^2 = \max_k G_k^2$. Rearranging the above inequality,
\begin{align*}
    &\mathbb{E}[\|\nabla F(\boldsymbol{w}^{\scriptstyle(t)})\|_2^2] \leq -\frac{2}{\eta_c^{\scriptstyle(t)}}\mathbb{E}[F(\tilde{\boldsymbol{w}}^{\scriptstyle(t + 1)}) - F(\tilde{\boldsymbol{w}}^{\scriptstyle(t)})]\\
    &\quad + \Biggl\{(\eta_c^{\scriptstyle(0)})^2\beta_c^2\left(\frac{(K - K_m)^2}{K_m^2} + \varepsilon\right)\frac{16(1 - \nu)}{\nu^2} E_c^3\\
    &\quad\qquad + (\eta_c^{\scriptstyle(t)})^2\beta_c^2\cdot\frac{2}{3}E_c^3 + \eta_c^{\scriptstyle(t)}\cdot 2\beta_c\frac{K}{K_m}E_c^2\Biggl\}G_{k, max}^2\\
    &\quad + \Biggl((\eta_s^{\scriptstyle(t - 1)})^2\cdot 4\varepsilon\beta_c^2 E_c E_s^2 + \frac{(\eta_s^{\scriptstyle(t - 1)})^2}{(\eta_c^{\scriptstyle(t)})^2}\cdot\frac{E_s^2}{E_c}\\
    &\quad\qquad + \frac{(\eta_s^{\scriptstyle(t - 1)})^2}{\eta_c^{\scriptstyle(t)}}\cdot 2\beta_c E_s^2\Biggl)G_s^2\\
    &\;\leq -\frac{2}{\eta_c^{\scriptstyle(T)}}\mathbb{E}[F(\tilde{\boldsymbol{w}}^{\scriptstyle(t + 1)}) - F(\tilde{\boldsymbol{w}}^{\scriptstyle(t)})]\\
    &\quad + \Biggl\{(\eta_c^{\scriptstyle(0)})^2\beta_c^2\left(\frac{(K - K_m)^2}{K_m^2} + \varepsilon\right)\frac{16(1 - \nu)}{\nu^2} E_c^3\\
    &\quad\qquad + (\eta_c^{\scriptstyle(0)})^2\beta_c^2\cdot\frac{2}{3}E_c^3 + \eta_c^{\scriptstyle(0)}\cdot 2\beta_c\frac{K}{K_m}E_c^2\Biggl\}G_{k, max}^2\\
    &\quad + \Biggl((\eta_s^{\scriptstyle(0)})^2\cdot 4\varepsilon\beta_c^2 E_c E_s^2 + \frac{(\eta_s^{\scriptstyle(0)})^2}{(\eta_c^{\scriptstyle(T)})^2}\cdot\frac{E_s^2}{E_c}\\
    &\quad\qquad + \frac{(\eta_s^{\scriptstyle(0)})^2}{\eta_c^{\scriptstyle(T)}}\cdot 2\beta_c E_s^2\Biggl)G_s^2,
\end{align*}
where the last inequality comes from the assumption that both $\eta_c^{\scriptstyle(t)}$ and $\eta_s^{\scriptstyle(t)}$ monotonically decrease w.r.t. $t$. Averaging the above inequality over iteration from $0$ to $T - 1$ and applying $\eta_c^{\scriptstyle(t)} = \alpha(t)/\sqrt{T}$ and $\eta_s^{\scriptstyle(t)} = \alpha(t)/T^{\frac{3}{4}}$, where $\alpha(t)$ is a monotonic decreasing function with an $\mathcal{O}(1)$ order, the first term in the right-hand size is upper-bounded as $\mathbb{E}[F(\tilde{\boldsymbol{w}}^{\scriptstyle(0)}) - F(\tilde{\boldsymbol{w}}^{\scriptstyle(T)})] \leq \mathbb{E}[F(\boldsymbol{w}^{\scriptstyle(0)})] - F(\boldsymbol{w}^*)$ since $\tilde{\boldsymbol{w}}^{\scriptstyle(0)} = \boldsymbol{w}^{\scriptstyle(0)}$ and $F(\boldsymbol{w}) \geq F(\boldsymbol{w}^*)$ for all $\boldsymbol{w}\in\mathbb{R}^N$. Then, we finally obtain the convergence of our proposed FL algorithm after averaging:
\begin{align*}
    &\mathbb{E}\left[\frac{1}{T}\sum_{t = 0}^{T - 1}\|\nabla F(\boldsymbol{w}^{\scriptstyle(t)})\|_2^2\right] \leq \frac{1}{\sqrt{T}}\\
    &\quad\times \left(\frac{2}{\alpha(T)}(\mathbb{E}[F(\boldsymbol{w}^{\scriptstyle(0)})] - F(\boldsymbol{w}^*)) + \mathrm{A} + \frac{1}{\sqrt{T}}\mathrm{B} + \frac{1}{T}\mathrm{C}\right),
\end{align*} where A, B, and C are given in \textbf{Theorem \ref{thm:convergence rate}}.

\vspace{-3mm}
\section{Proof of Lemma \ref{lm:upper bound of (a.2)}}\label{apdx:upper bound of (a.2)}
Using the simple inequality such as $\|\mathbf{a} - \mathbf{c}\|_2^2 \leq 2(\|\mathbf{a} - \mathbf{b}\|_2^2 + \|\mathbf{b} - \mathbf{c}\|_2^2)$, $(a.2)$ in Appendix \ref{apdx:convergence rate} is upper bounded as
\begin{align*}
    (a.2) &= \sum_{e = 0}^{E_c - 1}\mathbb{E}\left[\left\|\nabla F(\tilde{\boldsymbol{w}}^{\scriptstyle(t)}) - \sum_{k\in\mathcal{A}}p_k\nabla F_k(\boldsymbol{w}_k^{\scriptstyle(t, e)})\right\|_2^2\right]\\
    &\leq 2\underbrace{\sum_{e = 0}^{E_c - 1}\mathbb{E}\left[\|\nabla F(\tilde{\boldsymbol{w}}^{\scriptstyle(t)}) - \nabla F(\boldsymbol{w}^{\scriptstyle(t)})\|_2^2\right]}_{(a.2.1)}\\
    &\quad + 2\underbrace{\sum_{e = 0}^{E_c - 1}\mathbb{E}\left[\left\|\nabla F(\boldsymbol{w}^{\scriptstyle(t)}) - \sum_{k\in\mathcal{A}}p_k\nabla F_k(\boldsymbol{w}_k^{\scriptstyle(t, e)})\right\|_2^2\right]}_{(a.2.2)}.
\end{align*}

Before upper-bounding $(a.2.1)$, \textbf{Lemma \ref{lm:error memory norm boundedness}} is introduced.
\begin{lemma}\label{lm:error memory norm boundedness}
Under \emph{\textbf{Assumption \ref{as:gradient norm boundedness}}} and \emph{\textbf{Lemma \ref{lm:client sampling unbiasedness}}}, the expected squared norm of the error memory is upper bounded such as
\begin{align*}
    \mathbb{E}[\|\mathbf{m}_k^{\scriptstyle(t + 1)}\|_2^2] \leq \frac{4(1 - \nu)}{\nu^2}(\eta_c^{\scriptstyle(0)})^2 E_c^2 G_k^2
\end{align*}
for all $k$ and $t$, where $0 < \nu \leq 1$.
\begin{IEEEproof}
    See \emph{Appendix \ref{apdx:error memory norm boundedness}}.
\end{IEEEproof}
\end{lemma}

We can then upper-bound $(a.2.1)$:
\begin{align*}
    &(a.2.1) \leq E_c\beta_c^2\mathbb{E}[\|\tilde{\boldsymbol{w}}^{\scriptstyle(t)} - \boldsymbol{w}^{\scriptstyle(t)}\|_2^2]\\
    &\quad\leq 2E_c\beta_c^2\left(\mathbb{E}\left[\left\|\frac{K - K_m}{K_m}\mathbf{u}\right\|_2^2\right] + \mathbb{E}[\|\mathbf{u} - \mathbf{v}\|_2^2]\right)\\
    &\quad\leq 2E_c\beta_c^2\Biggl\{\frac{(K - K_m)^2}{K_m^2}\mathbb{E}\left[\sum_{k\in\mathcal{A}}p_k\|\mathbf{m}_k^{\scriptstyle(t)}\|_2^2\right]\\
    &\qquad\qquad\qquad + \varepsilon\Biggl(\mathbb{E}\left[\sum_{k\in\mathcal{A}}p_k\|\mathbf{m}_k^{\scriptstyle(t)}\|_2^2\right] + (\eta_s^{\scriptstyle(t - 1)})^2 E_s\\
    &\qquad\qquad\qquad\qquad \times \sum_{e = 0}^{E_s - 1}\mathbb{E}[\|\nabla F_s^{\scriptstyle(t - \frac{1}{2}, e)}(\boldsymbol{w}_s^{\scriptstyle(t - \frac{1}{2}, e)})\|_2^2]\Biggl)\Biggl\}\\
    &\quad\leq 2E_c\beta_c^2 \times \Biggl\{\varepsilon(\eta_s^{\scriptstyle(t - 1)})^2 E_s^2 G_s^2\\
    &\qquad + \left(\frac{(K - K_m)^2}{K_m^2} + \varepsilon\right)\frac{4(1 - \nu)}{\nu^2}(\eta_c^{\scriptstyle(0)})^2 E_c^2\sum_{k\in\mathcal{A}}p_k G_k^2\Biggl\},
\end{align*}  
where the first inequality is due to \textbf{Assumption \ref{as:smoothness}}; the second inequality is obtained by a simple inequality such as $\|\mathbf{a} + \mathbf{b}\|_2^2 \leq 2(\|\mathbf{a}\|_2^2 + \|\mathbf{b}\|_2^2)$ and letting $\mathbf{u} = \sum_{k\in\mathcal{A}}p_k\mathbf{m}_k^{\scriptstyle(t)}$ and $\mathbf{v} = \eta_s^{\scriptstyle(t - 1)}\sum_{e = 0}^{E_s - 1}\nabla F_s^{\scriptstyle(t - \frac{1}{2}, e)}(\boldsymbol{w}_s^{\scriptstyle(t - \frac{1}{2}, e)})$; the third inequality comes from \textbf{Proposition \ref{pro:error memory compensation}} and Jensen's inequality; the last inequality comes from \textbf{Lemma \ref{lm:error memory norm boundedness}} and \textbf{Assumption \ref{as:gradient norm boundedness}}. Also, under \textbf{Assumption \ref{as:smoothness}}, we can upper-bound $(a.2.2)$: 
\begin{align*}
    (a.2.2) &= \sum_{e = 0}^{E_c - 1}\mathbb{E}\left[\left\|\sum_{k\in\mathcal{A}}p_k(\nabla F_k(\boldsymbol{w}^{\scriptstyle(t)}) - \nabla F_k(\boldsymbol{w}_k^{\scriptstyle(t, e)}))\right\|_2^2\right]\\
    &\leq \beta_c^2\sum_{k\in\mathcal{A}}p_k\sum_{e = 0}^{E_c - 1}\mathbb{E}[\|\boldsymbol{w}_k^{\scriptstyle(t, 0)} - \boldsymbol{w}_k^{\scriptstyle(t, e)}\|_2^2]\\
    &= \beta_c^2\sum_{k\in\mathcal{A}}p_k\sum_{e = 0}^{E_c - 1}\mathbb{E}\left[\left\|\sum_{i = 0}^{e - 1}\eta_c^{\scriptstyle(t)}\nabla F_k^{\scriptstyle(t, i)}(\boldsymbol{w}_k^{\scriptstyle(t, i)})\right\|_2^2\right]\\
    &\leq \beta_c^2(\eta_c^{\scriptstyle(t)})^2\sum_{k\in\mathcal{A}}p_k\sum_{e = 0}^{E_c - 1}e^2 G_k^2\\
    &= \frac{E_c(E_c - 1)(2E_c - 1)}{6}\beta_c^2(\eta_c^{\scriptstyle(t)})^2\sum_{k\in\mathcal{A}}p_k G_k^2\\
    &\leq \frac{1}{3}E_c^3\beta_c^2(\eta_c^{\scriptstyle(t)})^2\sum_{k\in\mathcal{A}}p_k G_k^2,
\end{align*}
where the first inequality comes from $\boldsymbol{w}^{\scriptstyle(t)} = \boldsymbol{w}_k^{\scriptstyle(t, e)}$ and Jensen's inequality; the second equality comes from the local iterations in \eqref{eq:local iteration}; the second inequality comes from Jensen's inequality and \textbf{Assumption \ref{as:gradient norm boundedness}}. Finally, using the above inequalities, we get the upper bound of $(a.2)$:
\begin{align*}
    &(a.2)\leq 4E_c\beta_c^2\\
    &\;\;\times \Biggl\{\left(\frac{(K - K_m)^2}{K_m^2} + \varepsilon\right)\frac{4(1 - \nu)}{\nu^2}(\eta_c^{\scriptstyle(0)})^2 E_c^2\sum_{k\in\mathcal{A}}p_k G_k^2\\
    &\quad + \varepsilon(\eta_s^{\scriptstyle(t - 1)})^2 E_s^2 G_s^2\Biggl\} + \frac{2}{3}E_c^3\beta_c^2(\eta_c^{\scriptstyle(t)})^2\sum_{k\in\mathcal{A}}p_k G_k^2.
\end{align*}

\vspace{-3mm}
\section{Proof of Lemma \ref{lm:error memory norm boundedness}}\label{apdx:error memory norm boundedness}
The proof in this section is similar to the proof of Lemma $3$ in \cite{karimireddy2019error}. We use Definition $2.1$ in \cite{stich2018sparsified} as an assumption\footnote{Most compression methods satisfy this assumption \cite{stich2018sparsified}.} for the $\mathsf{Compress}(\cdot)$ in this paper:
\begin{align}\label{eq:contraction}
    \mathbb{E}[\|\mathbf{x} - \mathsf{Compress}(\mathbf{x})\|_2^2] \leq (1 - \nu)\|\mathbf{x}\|_2^2,
\end{align}
where $\mathbf{x}\in\mathbb{R}^{D}$ and $0 < \nu \leq 1$.

Moreover, before proceeding with the proof, we need the following two inequalities of \eqref{eq:young} and \eqref{eq:jensen and gradient norm boundedness}. Using AM-GM inequality, $\|\mathbf{a} + \mathbf{b}\|_2^2$ is upper-bounded:
\begin{align}\label{eq:young}
    \|\mathbf{a} + \mathbf{b}\|_2^2 &= \|\mathbf{a}\|_2^2 + \|\mathbf{b}\|_2^2 + 2\mathbf{a}^\textsf{T}\mathbf{b}\nonumber\\
    &\leq \|\mathbf{a}\|_2^2 + \|\mathbf{b}\|_2^2 + \gamma\|\mathbf{a}\|_2^2 + \frac{1}{\gamma}\|\mathbf{b}\|_2^2\nonumber\\
    &= (1 + \gamma)\|\mathbf{a}\|_2^2 + \left(1 + \frac{1}{\gamma}\right)\|\mathbf{b}\|_2^2.  
\end{align}
Using Jensen's inequality and \textbf{Assumption \ref{as:gradient norm boundedness}},
\begin{align}\label{eq:jensen and gradient norm boundedness}
    \mathbb{E}\left[\left\|\sum_{e = 0}^{E_c - 1}\nabla F_k^{\scriptstyle(t, e)}(\boldsymbol{w}_k^{\scriptstyle(t, e)})\right\|_2^2\right] \leq E_c^2 G_k^2.
\end{align}

Then, starting with \eqref{eq:error memory update} and assuming that $\eta_c^{\scriptstyle(t)}$ is monotonically decreasing w.r.t. $t$,
\begin{align} \label{eq:ineq_lemma3}
    &\mathbb{E}[\|\mathbf{m}_k^{\scriptstyle(t + 1)}\|_2^2]\nonumber\\
    &\quad\leq (1 - \nu)\mathbb{E}\left[\left\|\mathbf{m}_k^{\scriptstyle(t)} + \eta_c^{\scriptstyle(t)}\sum_{e = 0}^{E_c - 1}\nabla F_k^{\scriptstyle(t, e)}(\boldsymbol{w}_k^{\scriptstyle(t, e)})\right\|_2^2\right]\nonumber\\
    &\quad\leq (1 - \nu)(1 + \gamma)\mathbb{E}[\|\mathbf{m}_k^{\scriptstyle(t)}\|_2^2]\nonumber\\
    &\qquad + (1 - \nu)\left(1 + \frac{1}{\gamma}\right)(\eta_c^{\scriptstyle(0)})^2 E_c^2 G_k^2,
\end{align}
where the first inequality comes from \eqref{eq:contraction} and the second inequality comes from \eqref{eq:young} and \eqref{eq:jensen and gradient norm boundedness}. 

Finally, using the inequality in \eqref{eq:ineq_lemma3} recursively,
\begin{align*}
    &\mathbb{E}[\|\mathbf{m}_k^{\scriptstyle(t + 1)}\|_2^2]\\
    &\quad\leq (\eta_c^{\scriptstyle(0)})^2 E_c^2 G_k^2 (1 - \nu)(1 + \frac{1}{\gamma})\sum_{i = 0}^\infty[(1 - \nu)(1 + \gamma)]^i\\
    &\quad=(\eta_c^{\scriptstyle(0)})^2 E_c^2 G_k^2\frac{(1 - \nu)(1 + \frac{1}{\gamma})}{1 - (1 - \nu)(1 + \gamma)}\\
    &\quad\leq \frac{4(1 - \nu)}{\nu^2}(\eta_c^{\scriptstyle(0)})^2 E_c^2 G_k^2,
\end{align*} where the last inequality is satisfied by setting $\gamma = \frac{\nu}{2(1 - \nu)}$, resulting in $1 + \frac{1}{\gamma} = \frac{2}{\nu} - 1 \leq \frac{2}{\nu}$.

%
\vspace{-3mm}
\bibliographystyle{IEEEtran}  
\bibliography{IEEEabrv,reference}

\end{document}